\documentclass[letterpaper,11pt]{article}
\usepackage{caption}
\pdfoutput=1 
\usepackage{xcolor}
\usepackage{enumerate}
\usepackage{framed}
\usepackage{mdframed}
\usepackage{amsmath}
\usepackage{amsfonts}
\usepackage{mathrsfs}
\usepackage{amssymb}
\usepackage[normalem]{ulem}
\usepackage{graphicx, rotating}
\usepackage{epsfig}
\usepackage{latexsym}
\usepackage{graphicx}
\usepackage{color}
\usepackage{amsmath,bm,amssymb}
\usepackage{cite}
\usepackage{slashed}
\usepackage{hyperref}
\usepackage{epstopdf}
\usepackage[title]{appendix}
\usepackage[font=footnotesize,labelfont=]{caption}
\AppendGraphicsExtensions{.tif}
\hypersetup{colorlinks=true, citecolor=bluscuro, linkcolor=black, urlcolor=bluscuro}
\definecolor{rossos}{cmyk}{0,1,1,0.55}
\definecolor{bluscuro}{rgb}{0.15, 0.2, .85}
\definecolor{bluchiaro}{cmyk}{1,.3,0.,0.1}
\definecolor{Green}{rgb}{0, 0.65, 0.31}

\newcommand{\be}{\begin{equation}}
\newcommand{\ee}{\end{equation}}
\newcommand{\bea}{\begin{eqnarray}}
\newcommand{\eea}{\end{eqnarray}}
\newcommand{\beas}{\begin{eqnarray*}}
\newcommand{\eeas}{\end{eqnarray*}}

\newcommand{\rd}{{\rm d}}
\newcommand{\vev}[1]{\left\langle #1 \right\rangle}

\newcommand{\lp}{\left (}
\newcommand{\rp}{\right )}

\begin{document}
\def\thefootnote{\fnsymbol{footnote}}

\begin{center}
\LARGE{\textbf{Gravitational memory and Ward identities \\ in the local detector frame}} \\[0.5cm]
 
\large{Valerio De Luca,$^{1}$\footnote{\href{mailto:}{\texttt{vdeluca@sas.upenn.edu}}}, Justin Khoury,$^{1}$\footnote{\href{mailto:}{\texttt{jkhoury@sas.upenn.edu}}} and Sam S. C. Wong,$^{2}$\footnote{\href{mailto:}{\texttt{samwong@cityu.edu.hk}}}}
\\[0.5cm]

\small{
\textit{
$^1$Center for Particle Cosmology, Department of Physics and Astronomy, \\ University of Pennsylvania, 209 South 33rd St, Philadelphia, PA 19104, USA\\
 \vskip 5pt
~~$^2$Department of Physics, City University of Hong Kong, \\
Tat Chee Avenue, Kowloon, Hong Kong SAR, China}
 }

\vspace{.2cm}

\end{center}

\vspace{.6cm}

\hrule \vspace{0.2cm}
\centerline{\small{\bf Abstract}}
%\vspace{-0.2cm}
{\small\noindent 
Gravitational memory, which describes the permanent shift in the strain after the passage of gravitational waves, 
is directly related to Weinberg's soft graviton theorems and the Bondi-Metzner-Sachs (BMS) symmetry group of
asymptotically flat space-times. In this work, we provide an equivalent description of the phenomenon in local coordinates around gravitational wave detectors, such as transverse-traceless (TT) gauge. We show that gravitational memory is encoded in large residual diffeomorphisms in this gauge, which include time-dependent anisotropic spatial rescalings, and prove their equivalence to BMS transformations when translated to TT gauge. We then derive the associated Ward identities and associated soft theorems, for both scattering amplitudes and equal-time (in-in) correlation functions, and explicitly check their validity for planar gravitational waves. The in-in identities are recognized as the flat-space analog of the well-known inflationary consistency relations.}

\vspace{0.3cm}
\noindent
\hrule
\def\thefootnote{\arabic{footnote}}
\setcounter{footnote}{0}

\tableofcontents

\section{Introduction}

In recent years, the detection of gravitational waves (GWs) from compact binaries has provided a new avenue to
test predictions of general relativity (GR), in the regime of strongly gravitating and rapidly evolving space-times, as well as to probe fundamental physics~\cite{LIGOScientific:2016lio,LIGOScientific:2018dkp, LIGOScientific:2020tif, LIGOScientific:2021sio}. The large number of events observed so far are consistent with GR within statistical uncertainties of the measurements. The improved sensitivity of next-generation detectors will allow us to probe the predictions of GR even more precisely, possibly accessing subdominant phenomena in the gravitational waveforms. Among these, one of the peculiar predictions of GR is the presence of nonoscillatory contributions to the GW strain, in addition to the familiar oscillatory terms. This persistent, nonoscillatory contribution is known as gravitational memory.

Gravitational memory usually refers to a lasting change in the GW strain that occurs for many types of transient GW sources. It is determined by freely falling observers who measure enduring changes in their separation before and after the GW burst. There are various types of memory effects. A linear component of the memory was first predicted by Zel'dovich and Polnarev in linearized gravity in the context of the gravitational scattering of compact objects~\cite{Zeldovich:1974gvh}. A nonlinear contribution from full GR was shown to arise from the GW energy flux and from the cumulative effect of GWs on the stress-energy tensor~\cite{Christodoulou:1991cr,Blanchet:1992br,Thorne:1992sdb}. (See Refs.~\cite{PhysRevD.44.R2945, Favata:2008yd, Blanchet:2020ngx, Blanchet:2023pce, Cunningham:2024dog} for their computation within the context of the multipolar-expanded post-Newtonian/post-Minkowskian approximations, and Refs.~\cite{Pollney:2010hs, Grant:2023jhd, Yoo:2023spi} for numerical relativity simulations.) These two types are usually referred to as {\it displacement memory}, arising for observers who are initially comoving. 

When the observers have an initial relative velocity, additional subleading memory effects can be measured, such as the spin and center of mass memories. Spin memory manifests itself as a relative time delay between two observers in counterorbiting trajectories. It originates from a nonvanishing change in the magnetic parity part of the time integral of the GW strain induced by fluxes of angular momentum per unit solid angle~\cite{Nichols:2017rqr, Flanagan:2018yzh}. The center of mass memory is related to changes in the center of mass part of the angular momentum of a space-time~\cite{Nichols:2018qac}. Both families of memory effects were shown to fall under the broader class of persistent observables, beyond the context of asymptotically flat space-times. A subset of these forms the curve deviation, which allows us to probe test masses with an initial relative acceleration~\cite{Flanagan:2018yzh,Flanagan:2019ezo,Grant:2021hga}.

On the experimental side, memory effects can be detected either from individual events or with population-based analyses. The detection channel depends sensitively on the experiment under consideration. For individual events, current ground-based detectors may detect nonlinear displacement memories from gravitationally bound systems only for sufficiently loud sources. To date, there is no evidence for memory effects in any of the individual detections by LIGO and Virgo~\cite{Favata:2009ii, Johnson:2018xly, Hubner:2021amk}, and it is unlikely that they will be detected even as detectors reach their design and ``plus'' sensitivities. On the other hand, space-based interferometers like DECIGO~\cite{Kawamura:2011zz,Kawamura:2020pcg}, LISA~\cite{LISA:2017pwj, LISA:2022kgy}, Taiji~\cite{Ruan:2018tsw}, and TianQin~\cite{TianQin:2015yph} may measure the displacement memory arising from mergers of stellar-mass or supermassive black-hole binaries~\cite{Favata:2010zu,Islo:2019qht,Gasparotto:2023fcg,Inchauspe:2024ibs, Hou:2024rgo}, while it will take longer for pulsar timing array experiments~\cite{Pshirkov:2009ak, NANOGrav:2019vto,Islo:2019qht, vanHaasteren:2022agf}. Finally, next-generation ground-based detectors such as the Einstein Telescope~\cite{Punturo:2010zz, Branchesi:2023mws} and Cosmic Explorer~\cite{Reitze:2019iox} are forecasted to be sensitive enough to measure the displacement memory from individual events~\cite{Johnson:2018xly}. In the category of population-driven analyses, the method of ``stacking'' combines many low-significance events to give a single higher-significance event that exceeds a threshold for memory detection~\cite{Lasky:2016knh}. For example, it was shown that hundreds to thousands of events would be necessary for the  current LIGO/Virgo sensitivities to reach a detection~\cite{Lasky:2016knh, Hubner:2019sly, Boersma:2020gxx, Hubner:2021amk}. However, in a few years of observations at their design sensitivities, or using next-generation detectors, both nonlinear and spin memory effects could be observed~\cite{Boersma:2020gxx,Grant:2022bla, Goncharov:2023woe}.

Memory effects have close connections to the BMS symmetry group of asymptotically flat space-times and its conserved charges~\cite{1962RSPSA.269...21B, Sachs:1962wk, Sachs:1962zza, Strominger:2017zoo}. In particular, the displacement memory is related to the supertranslation symmetries and charges, which provide a superset of the known Poincar\'e group.
The displacement memory arises as the permanent shift of the asymptotic shear after the passage of GWs, and it can be equivalently described as a transition between two different asymptotic BMS frames related by a supertranslation~\cite{Strominger:2013jfa,Ashtekar:2014zsa,Flanagan:2015pxa, Kehagias:2016zry}.
Similarly, the extended BMS group, which includes superrotation symmetries corresponding to supermomentum and superspin charges, is related to the subleading 
spin and center of mass memory effects~\cite{Barnich:2009se, Barnich:2010eb, Kapec:2014opa, Nichols:2018qac, Strominger:2017zoo, Himwich:2019qmj}. Additional BMS symmetry groups were later proposed~\cite{Campiglia:2015yka, Freidel:2021fxf, Godazgar:2018qpq}, resulting in additional memory-type effects~\cite{Compere:2018ylh, Seraj:2021rxd, Godazgar:2022pbx, Siddhant:2024nft}.

Memory effects and asymptotic symmetries represent two corners of the so-called ``infrared triangle''~\cite{Strominger:2017zoo}, which establishes universal relationships between them and Weinberg's soft graviton theorems in quantum field theory~\cite{Weinberg:1965nx}. Each corner of the triangle is an equivalent way of characterizing gravitational physics at large distances. In this context, it has been shown that the displacement memory is connected to the soft graviton theorem~\cite{Strominger:2013jfa, He:2014laa, Strominger:2014pwa, Mirbabayi:2016xvc}, while the spin memory is associated with the subleading soft graviton theorem~\cite{Strominger:2017zoo,Anupam:2018vyu, Himwich:2019qmj}.
The fundamental relation between soft theorems and asymptotic symmetries is ubiquitous in physics, and it has been extensively discussed also in the context of  cosmology~\cite{Creminelli:2012ed,Kehagias:2013yd, Peloso:2013zw, Hinterbichler:2013dpa, Hui:2018cag, Hui:2022dnm}.

The relation between BMS transformations and the description of gravitational memory in the more familiar ``local" coordinate system of GW detectors is, however, not immediately clear. Freely falling detectors like LISA are usually described in terms of synchronous or TT coordinates. This suggests an equivalent description of the associated BMS asymptotic symmetries in the TT frame. A goal of this work is to elucidate this relation. This paper is a more detailed companion to a short paper~\cite{DeLuca:2024bpt}, which summarizes the salient points.

We first determine the residual coordinate transformations in the local TT frame around the detector that capture the physics of the gravitational memory effect. We will find that gravitational memory corresponds to large residual diffeomorphisms in this gauge, such as anisotropic (volume-preserving) spatial rescalings. As we will show, these diffeomorphisms are precisely equivalent to BMS transformations when translated to the local TT coordinates. We then derive the consistency relations/soft theorems, both for in-in correlators as well as for scattering amplitudes, starting from the Ward identities associated with the residual diffeomorphisms. Remarkably, the resulting soft theorems are the flat-space analog of the inflationary consistency relations with a soft tensor mode.

\section{Gravitational memory and BMS symmetry}

In this Section we review the connection between gravitational memory and BMS transformations.  We first discuss the radiative/Bondi coordinates, often employed in the context of gravitational radiative modes. We then provide the relation between this gauge and TT coordinates, usually considered in the context of GW observations. In doing so, we follow~\cite{Blanchet:2020ngx} and establish how memory enters the GW strain. This will allow us to show how a BMS transformation can be used to describe gravitational memory~\cite{Strominger:2013jfa}.

Throughout this work, we use natural units~$\hbar = c = 1$, and a mostly positive signature for the metric. Radiative coordinates, centered at the location of the source of the gravitational radiation, are denoted with capital letters as~$X^\mu = (T,R, \theta^a)$, with~$a,b, \cdots = \{1,2\}$ denoting coordinates on the two-sphere, and $i,j, \cdots =
\{1, 2, 3\}$ denoting Cartesian spatial indices, raised and lowered with the Kronecker metric $\delta_{ij}$.  We also introduce the metric tensor~$g_{\mu \nu}$ and the retarded time $U = T- R$.

\subsection{Radiative coordinates}
\label{Bondi gauge}

The class of {\it radiative coordinates}, such as Bondi~\cite{Bondi:1962px,Sachs:1962wk} and Newman-Unti gauge, are frequently used to study asymptotically flat space-times characterized by the presence of outgoing radiation generated by an isolated matter system. Technically, to be well defined, this class of coordinates requires the system to be stationary before some finite time in the past, such that its radiative multipole moments (defined below) are constant before that time.

Radiative coordinates are defined by the gauge condition~\cite{Barnich:2009se, Barnich:2010eb, Strominger:2013jfa, Strominger:2017zoo,Grant:2021hga}
\begin{equation}
g_{RR} = 0\,; \qquad g_{Ra} = 0\,; \qquad \det [g_{ab}] = R^4 q(\theta^a)\,,
\label{Bondi coordinate condition}
\end{equation}
where~$q$ depends only on angular coordinates. In Bondi gauge, for instance, the metric takes the general form~\cite{Barnich:2009se, Barnich:2010eb, Strominger:2013jfa, Strominger:2017zoo,Grant:2021hga}
\begin{align}
{\rm d}s^2 = - \left( 1 - \frac{2 \mathbb{M}}{R} \right) {\rm e}^{2\beta/R} {\rm d} U^2 - 2 {\rm e}^{2\beta/R} {\rm d} U  {\rm d} R + R^2 \mathcal{H}_{ab} \left( {\rm d} \theta^a - \frac{\mathcal{W}^a}{R^2} {\rm d} U \right) \left( {\rm d} \theta^b - \frac{\mathcal{W}^b}{R^2} {\rm d} U \right)\,.
\end{align}
The metric functions~$\beta$,~$\mathbb{M}$,~$\mathcal{W}^a$, and~$\mathcal{H}_{ab}$ are determined by solving Einstein’s field equations, subject to initial data on a null hypersurface (see Ref.~\cite{Madler:2016xju} for further details). To completely specify the metric components, one must also impose boundary conditions at null infinity, sending $R \to \infty$ and keeping $U$ fixed, which constrain their falloff behavior:
\begin{equation}
\lim_{R \to \infty} \frac{\beta}{R} = \lim_{R \to \infty} \frac{\mathcal{W}^a}{R} = \lim_{R \to \infty} \frac{\mathbb{M}}{R} = 0\,; \qquad \lim_{R \to \infty} \mathcal{H}_{ab} = \gamma_{ab}\,,
\end{equation}
where~$\gamma_{ab}$ is the standard metric on the two-sphere. To highlight the leading~$1/R$ behavior and match to these boundary conditions, it is useful to rewrite the metric functions in the form~\cite{Barnich:2009se, Barnich:2010eb, Strominger:2013jfa, Strominger:2017zoo,Grant:2021hga}
\begin{align}
\beta & = \frac{1}{R} \tilde \beta (U,R, \theta^a)\,; \nonumber \\
\mathbb{M} & = m (U,\theta^a) + \frac{1}{R} \mathcal{M} (U,R,\theta^a)\,; \nonumber \\
\mathcal{W}^a & = W^a (U, \theta^b) + \frac{1}{R} V^a (U, \theta^b) + \frac{1}{R^2} \Upsilon^a (U,R, \theta^b)\,; \nonumber \\
\mathcal{H}_{ab} & = \sqrt{1 + \frac{\mathcal{C}_{cd}(U,R, \theta^e) \mathcal{C}^{cd}(U,R, \theta^e)}{2 R^2}} \gamma_{ab} + \frac{1}{R} \mathcal{C}_{ab} (U,R, \theta^c)\,,
\label{BCBMS}
\end{align}
in terms of~$m(U,\theta^a)$, known as the Bondi mass aspect, and the vectors~$W^a(U, \theta^b)$ and~$V^a(U, \theta^b)$. The remaining components,~$\tilde \beta$,~$\mathcal{M}$,~$\Upsilon^a$ and the symmetric trace-free tensor~$\mathcal{C}_{ab}$, depend on all four space-time coordinates and admit an expansion in powers of $1/R$, starting at $\mathcal{O}(1)$. For instance,~$\mathcal{C}_{ab}$ can be expanded as  
\begin{equation}
\mathcal{C}_{ab} (U,R, \theta^c) = C_{ab} (U,\theta^c) + \sum_{n = 2}^\infty \frac{1}{R^n} \mathcal{E}^{(n)}_{ab} (U, \theta^c)\,,
\end{equation}
where the leading term,~$C_{ab}$, is the shear, and the subleading coefficients are the higher Bondi aspects $\mathcal{E}^{(n)}_{ab}$~\cite{Compere:2022zdz}.
The retarded time derivative of the shear is usually denoted as the news tensor,
\be
N_{ab} = \partial_U C_{ab}\,. 
\ee
Lastly, the vector~$V^a$ can be expressed in terms of the angular momentum aspect~$N^a$, defined as
\begin{equation}
N^a (U, \theta^b)  = - \frac{3}{2} V^a + \frac{3}{32} D^a (C_{bc} C^{bc}) + \frac{3}{4} C^{ab} D^c C_{bc}\,,
\end{equation}
where~$D_a$ is the covariant derivative on the two-sphere. In the following, we will be primarily interested in the asymptotic future null infinity limit, keeping the leading~$1/R$ terms of the metric functions, with~$U$ fixed.

The explicit form of the metric functions can be obtained once the mass and angular momentum aspects, the shear and the news tensor are specified. To do so requires solving the linearized field equations outside the matter source, without any incoming flux from past null infinity. A standard approach is the multipolar post-Minkowskian (PM) approximation~\cite{Blanchet:2020ngx}, in which the metric perturbation is expanded in powers of~$G$. At leading order, the Bondi functions are given by~\cite{Blanchet:2020ngx,Blanchet:2023pce} 
\begin{align}
m & = \sum_{\ell = 0}^\infty \frac{(\ell+1)(\ell+2)}{2\ell!} N_{L} {\cal U}_{L}(U)\,; \nonumber \\
N^a & = e^i_{a}  \sum_{\ell = 1}^\infty \frac{(\ell+1)(\ell+2)}{2(\ell-1)!} N_{L-1} \left\{  {\cal U}_{i L-1}(U) +  N^k \epsilon_{rki} {\cal V}_{rL-1}(U)\right\}\,; \nonumber \\
   C_{ab} & = 4 e^i_{\langle a} e^j_{b\rangle} \sum_{\ell = 2}^\infty \frac{1}{\ell!} N_{L-2}  \left\{{\cal U}_{ij L-2}(U)  +  N^k \epsilon_{rk(i} {\cal V}_{j)rL-2}(U)\right\}\,,
\label{Bondi U and V}
\end{align}
where~${\cal U}_L$ and~${\cal V}_L$ are symmetric, trace-free (STF) Cartesian tensors,\footnote{The multi-index~$L$ stands for~$\ell$ symmetrized indices, {\it e.g.}, ${\cal U}_L = {\cal U}_{(i_1\cdots i_\ell)}$ and~$N_L = N_{i_1}\cdots N_{i_\ell}$. A summation over the repeated multi-index~$L$ is understood.} known as the radiative multipole moments. We will have more to say about them in Sec.~\ref{TT gauge} below. The unit vector~$N_i = X_i/R$ is normal to the sphere and points from the source to the observer. We have also introduced the components $e^i_a = \partial N^i/\partial \theta^a$ of the basis on the two-sphere, with $e^i_{\langle a} e^j_{b\rangle} = e^i_{(a} e^j_{b)} - \frac{1}{2} \gamma_{ab}P^{ij}$, where~$P_{ij} = \delta_{ij} - N_i N_j$ is the projector onto the sphere. Throughout this paper, we will often make use of the following identities~\cite{Blanchet:2020ngx}:
\be
N^i e^a_i = 0\,; \qquad \partial_i \theta^a = \frac{1}{R} \gamma^{ab}e^i_b\,; \qquad   \delta_{ij} e^i_ae^j_b = \gamma_{ab}\,; \qquad 
\gamma^{ab} e^i_a e^j_b= P_{ij}\,.
\label{identities}
\ee

By performing a suitable gauge transformation from radiative to TT coordinates, we will see below that the shear field~$C_{ab}$ contains the information of the gravitational strain and memory effects.

\subsection{Memory effects in TT coordinates}
\label{TT gauge}

When dealing with GW detection experiments, a convenient coordinate system to adopt is synchronous or TT frame, defined by~$g_{00} = 1$ and $g_{0i}= 0$.
Its convenience stems from the fact that freely falling mirrors, initially at rest, remain at fixed coordinates during the passage of GWs. Their proper separation is,
of course, affected by GWs, and is encoded, for instance, in the proper time taken by photons to travel along the interferometer arms.

Following the approach discussed in Ref.~\cite{Blanchet:2020ngx}, it is possible to draw a one-to-one correspondence between the GW strain~$H_{ij}^\text{\tiny TT}$ in TT gauge and the shear field~$C_{ab}$ of radiative Bondi coordinates, valid to leading order in~$1/R$:
\be
\frac{G}{R} C_{ab} = e^i_{\langle a} e^j_{b\rangle} H_{ij}^\text{\tiny TT}\,.
\label{shear HijTT}
\ee
Thus, from Eq.~\eqref{Bondi U and V}, the GW strain at leading order in~$1/R$ reads\footnote{Notice that Ref.~\cite{Blanchet:2020ngx} defines~$H_{ij}^\text{\tiny TT}$ without the~$G/R$ factor.}
\begin{align}
\label{HijPN}
H_{ij}^\text{\tiny TT} = \frac{4G}{R} \Pi_{ijkl}(N) \sum_{\ell=2}^\infty \frac{N_{L-2}}{\ell!} \Big\{  \, \mathcal{U}_{kl L-2} (U)  +  N^{m} \epsilon_{nm(k} \mathcal{V}_{l)nL-2} (U) \Big\}\,,
\end{align}
where we have introduced the TT projector
\be
\Pi_{ijkl} =  \frac{1}{2} \Big(P_{ik} P_{jl} + P_{il} P_{jk} -  P_{ij}P_{kl}\Big)\,.
\label{TT projector}
\ee
To understand the origin of gravitational memory, let us briefly comment on the different contributions to the radiative multipole moments~${\cal U}_L$ and~${\cal V}_L$. In general, these are fixed in terms of the  mass and current multipole moments,~$M_L$ and~$S_L$, of the source, by performing a post-Newtonian (PN) expansion and matching the solutions of the PN and PM approaches in a buffer region outside the object~\cite{Maggiore:2007ulw}. The first few terms resulting from this procedure are given by
\begin{align}
\nonumber
\mathcal{U}_{L} (U) &= M^{(\ell)}_{L}(U) + \mathcal{U}_{L}^\text{\tiny tail} (U) + \mathcal{U}_{L}^\text{\tiny memory} (U) + \ldots \\
\mathcal{V}_{L} (U) &= S^{(\ell)}_{L}(U) + \mathcal{V}_{L}^\text{\tiny tail} (U) + \ldots
\label{UV tail mem}
\end{align}
The leading terms are the instantaneous contributions, where superscripts indicate time derivatives, {\it e.g.},~$M^{(j)}_{L}(U) = \frac{\partial^j}{\partial U^j}M_{L}(U)$. For the quadrupole moment ($\ell = 2$), for instance, these give~$\mathcal{U}_{ij} = \ddot{M}_{ij}$ and~$\mathcal{V}_{ij} = \ddot{S}_{ij}$, such that~$H_{ij}^\text{\tiny TT} = \frac{2G}{R} \Pi_{ijkl}(N)  \left \{ \ddot{M}_{kl} + N_m \epsilon_{nm(k} \ddot{S}_{l)n} \right \}$. 
This is the familiar result that~$H_{ij}^\text{\tiny TT}$ is sourced by the second time derivative of the source quadrupole moments.  

The higher-order terms in Eq.~\eqref{UV tail mem} are more intricate and include hereditary contributions~\cite{Blanchet:1992br}.
The terms~$\mathcal{U}_{L}^\text{\tiny tail}$ and~$\mathcal{V}_{L}^\text{\tiny tail}$ are the leading tail effects, which arise from GWs scattering off the gravitational potential of the system with total mass $M$.
Explicitly,
\begin{align}
\nonumber
\mathcal{U}_{L}^\text{\tiny tail} (U) &=   2 G M \int_{-\infty}^U {\rm d} \tau  \left[ \log \left( \frac{U-\tau}{2 r_0} \right) + \kappa_L \right] M_{L}^{(\ell + 2)} (\tau)\,; \\
\mathcal{V}_{L}^\text{\tiny tail} (U) &= 2 G M \int_{-\infty}^U {\rm d} \tau  \left[ \log \left( \frac{U-\tau}{2 r_0} \right) + \pi_L \right] S_{L}^{(\ell + 2)} (\tau)\,,
\end{align}
where~$\kappa_L$ and~$\pi_L$ are constants,\footnote{See Eq.~(2.25) of Ref.~\cite{Favata:2008yd} for their explicit expressions.} and $r_0$ is an arbitrary timescale that disappears in physical observables. 
The term of interest in Eq.~\eqref{UV tail mem} for our purposes is the~$\mathcal{U}_{L}^\text{\tiny memory}$ contribution to the mass moments, which encodes the {\it nonlinear gravitational memory effect}. At the nonlinear level, the memory effect is sourced by the energy flux~${\rm d}^2E_\text{\tiny GW}/{\rm d} \Omega {\rm d} U$ of radiated GWs. This gives rise to a 2.5~PN contribution in the radiative mass moments of the form~\cite{Blanchet:1992br,Favata:2008yd}
\begin{equation}
\mathcal{U}_{L}^\text{\tiny memory} (U) =  \frac{2(2\ell + 1)!!}{(\ell + 1)(\ell + 2)}  \int_{-\infty}^{U} {\rm d} U' \int {\rm d} \Omega  \frac{{\rm d}^2E_\text{\tiny GW}}{{\rm d} \Omega {\rm d} U'} n_{\langle L \rangle} \,,
\label{UL nonlin memory}
\end{equation}
where~$n$ is a unit vector normal to the sphere at~$(\theta,\phi)$, and its subscript~$\langle L \rangle$ indicates STF projection.\footnote{Note that~$n$ should not be confused with~$N_i$, which points from the source to the observer.} The time integral gives exactly the hereditary nature to the memory term, since the GW field depends on the entire past history of the source. 

Focusing on~$\ell = 2$ for concreteness, substitution of Eq.~\eqref{UL nonlin memory} into Eq.~\eqref{HijPN} gives the following TT tensor field contribution:
\begin{equation}
H_{ij}^\text{\tiny TT} = \frac{4G}{R} \int_{-\infty}^{U} {\rm d}U' \left[ \int \frac{{\rm d}^2 E_\text{\tiny GW}}{{\rm d} \Omega {\rm d} U'}  \frac{n_i n_j}{1- \vec{n} \cdot \vec{N}} {\rm d} \Omega \right]^\text{\tiny TT}\,.
\label{Hij nonlinear memory}
\end{equation}
This emphatically shows that memory effects describe changes in the~$1/R$ spatial component of the TT projection of the metric perturbation, and not just of the~$1/R$ expansion (which might be confused with a change of the Coulomb potential)~\cite{Favata:2010zu}.\footnote{Equations~\eqref{UL nonlin memory} and~\eqref{Hij nonlinear memory} correspond to the nonlinear or Christodoulou gravitational memory. In general, there can also be a linear memory contribution. For instance, for a system of $N_g$ gravitationally unbound particles with masses $M_a$ and constant velocities $v_a$, a change in the derivatives of the mass multipoles due to hyperbolic encounters, $\Delta M^{(2)}_{ij}$, gives rise to the linear memory effect~\cite{Thorne:1992sdb}
\begin{equation}
H_{ij}^\text{\tiny TT} = \Delta \sum_{a=1}^{N_g} \frac{4 G M_a}{R\sqrt{1-v_a^2}} \left[ \frac{v_a^i v_a^j}{1 - \vec{v}_a \cdot \vec{N}} \right]^\text{\tiny TT}\,,
\end{equation}
in terms of the difference $\Delta$ between the late- and early-time values. The linear memory contribution generally vanishes for gravitationally bound systems, though see~\cite{Favata:2008yd} for exceptions. See also Ref.~\cite{Hait:2022ukn}
for related estimates.} To complete our discussion of Eq.~\eqref{UV tail mem}, let us mention that the ellipses include subleading tail terms (so-called tail-of-tail effects), as well as subleading instantaneous contributions. 

The direct relation between the shear and TT tensor field embodied in Eq.~\eqref{shear HijTT} shows that a change in the GW strain~$\Delta H_{ij}^\text{\tiny TT}$ induced by gravitational memory translates into a change of the shear $\Delta C_{ab}$ in Bondi coordinates. In the next Section we will dive deeper into the connection between BMS asymptotic symmetries and memory effects~\cite{Strominger:2013jfa}.

\subsection{BMS asymptotic symmetries}
\label{BMS rad}

BMS symmetries are diffeomorphisms acting on future null infinity that preserve its intrinsic geometric structure~\cite{Wald:1984rg, Geroch:1977big}. 
The corresponding infinitesimal symmetries are described by the vector field ${\vec \xi} =\xi^U \partial_U+ \xi^a \partial_a$, with components
\begin{align}
\label{BMSdiff}
\nonumber
\xi^U  & \equiv f(U, \theta^a) = T(\theta^a) + \frac{U}{2}  D_aY^a(\theta^b)\,; \\
\xi^a  & = Y^a (\theta^b)\,.
\end{align}
The scalar function~$T(\theta^a)$ generates the so-called supertranslation. The vector field~$Y^a$ satisfies
the conformal Killing equation on the two-sphere,~$2 D_{(a} Y_{b)} - D_c Y^c \gamma_{ab}=0$. In the standard BMS group, one
focuses on solutions to the conformal Killing equation which are everywhere smooth on the sphere, and as such form the~$SL(2,\mathbb{C})$ algebra (isomorphic to the Lorentz transformations). In extensions of the BMS group, one includes all local conformal Killing vectors and diffs on the two-sphere~\cite{Barnich:2009se}.

One can extend these diffeomorphims from future null infinity to the whole space-time by requiring that they maintain the retarded Bondi gauge condition~\eqref{Bondi coordinate condition} and the~$1/R$ scaling behavior of the metric components given in Eq.~\eqref{BCBMS}. The resulting generalized diffeomorphism is given by~${\vec \xi} =\xi^U \partial_U+  \xi^R \partial_R + \xi^a \partial_a$, with components\footnote{Here,~$D^2 = \gamma^{ab} D_a D_b$ is the Laplacian on the two-sphere.} 
\begin{align}
\label{BMSdiffgeneral}
\nonumber
\xi^U  & = f\,;\\ 
\nonumber
\xi^R  & = -\frac{R}{2} D_aY^a + \frac{1}{2} D^2 f + \mathcal{O} \left( \frac{1}{R} \right)\,;\\ 
\xi^a & = Y^a - \frac{1}{R} D^a f + \mathcal{O} \left( \frac{1}{R^2} \right)\,.
\end{align}
Under this coordinate transformation, the Bondi mass aspect, shear field and angular momentum aspect transform,  respectively, as~\cite{Flanagan:2015pxa,Nichols:2018qac}
\begin{align}
\delta m &= f \partial_U m + \frac{1}{4} N^{ab} D_aD_b f
+ \frac{1}{2} D_af D_b N^{ab}
+ \frac{3}{2} m D_aY^a + Y^a D_am
+ \frac{1}{8} C^{ab} D_aD_b D_e Y^e\,; \nonumber  \\
\delta C_{ab} &= f N_{ab} - 2 D_aD_b f + \gamma_{ab} D^2 f- \frac{1}{2} D_cY^c C_{ab} + {\cal L}_{Y} C_{ab}\,; \nonumber \\
\delta N_a &=  \big(f\partial_U +\mathcal L_Y + D_aY^a\big) N_a +
3m D_af - \frac{3}{4} D_b f\big(D^b D^c C_{ca} - D_aD_c C^{bc}\big)  + \frac 34 C_{ab} N^{bc} D_c f \,,
\end{align}
where~${\cal L}_Y$ is a Lie derivative with respect to~$Y^a$. In deriving these expressions we have assumed that the initial and final states are in vacuum.

Consider the displacement memory effect, which is relevant for initially comoving, freely falling adjacent observers.
This effect can be described as the change in the shear tensor,~$ \Delta C_{ab}$, associated with a transition between one canonical, nonradiative frame, where $C_{ab} = 0$, to a final, nonradiative region, where an intermediate burst of GWs has occurred. This transition is then interpreted as a BMS transformation relating the two frames, with
\begin{equation}
 \Delta C_{ab} =  - 2 D_aD_b f + \gamma_{ab} D^2 f\,,
\label{Delta C in terms of f}
\end{equation}
where we have used the property that~$C_{ab}=0$ in the canonical frame.
Thus the permanent shift of the asymptotic shear can be equivalently characterized as a transition between two different asymptotic BMS frames related by a supertranslation.

Expanding the time diffeomorphism~$f(U, \theta^a)$ in a multipolar STF decomposition~\cite{Blanchet:2020ngx,Blanchet:2023pce},
\be
f(U, \theta^a) = \sum_{\ell=0}^{\infty} N_L f_L(U)\,,
\ee
Eq.~\eqref{Delta C in terms of f} becomes
\begin{equation}
    \Delta C_{ab} = e^i_{\langle a} e^j_{b\rangle}  \sum_{\ell =2 }^{\infty} \ell (\ell-1) N_{L-2} f_{ij L-2}(U)\,.
\end{equation}
Compare this result with the expansion of~$C_{ab}$ in terms of~${\cal U}_{L}(U)$ and~${\cal V}_{L}(U)$ given in Eq.~\eqref{Bondi U and V}.
Since~$f(U, \theta^a)$ is at most linear in~$U$ according to Eq.~\eqref{BMSdiff}, we see that a BMS transformation
accounts for the constant and linear-in-$U$ terms in the radiative multipole moments. 
In other words,
\begin{eqnarray}
\label{fUV}
     f_{ij L-2}(U) = \frac{4}{\ell (\ell-1) \ell!}  {\cal U}_{ij L-2}^\text{\tiny lin}(U) \,,
\end{eqnarray}
where the ``lin" superscript indicates terms at most linear in~$U$. Notice that it is not possible to fix the odd parity term $\cal V$, since the BMS function $f = T + \frac{U}{2}D_a Y^a$ is built out of a scalar and the divergence of a vector. 
We will come back to this point  below.

\section{Residual diffeomorphisms in TT coordinates}
\label{sec: diffTT}

The procedure outlined in the previous Section establishes the relation between BMS transformations in Bondi coordinates and gravitational memory. 
The relation between BMS transformations and the description of gravitational memory in the more familiar ``local" coordinate system of GW detectors
is, however, not immediately clear. The purpose of this Section is to elucidate the form of the most general residual diffeomorphisms one can
perform in TT gauge to describe the memory effect. We will see how these diffeomorphisms are precisely equivalent to BMS transformations
when translated to the local TT coordinates.

For concreteness, consider the physical case of a binary system emitting GWs far enough from an observer,  which can be represented by a ground-based or space-based interferometer. This is the most relevant scenario for GW observations measured so far by the LIGO/Virgo/Kagra Collaboration. We then consider a coordinate frame centered at the location of the source, such that the observer is located at a radial distance $R \gg 1$. This hierarchy of scales allows us to focus on the leading $1/R$ contribution in the GW strain, similar to what is assumed in the radiative coordinates in the limit $R \to \infty$. 

\begin{figure}[t!]
    \centering
    \includegraphics[scale=0.6]{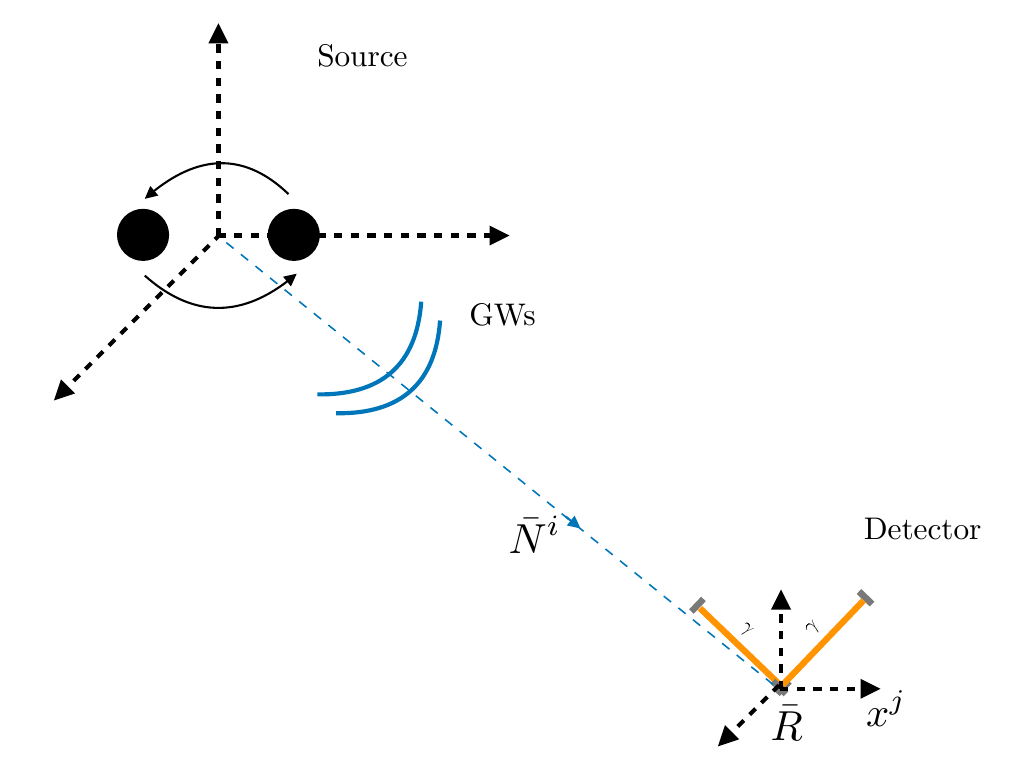}
    \caption{Schematic illustration of the coordinate systems around the source and, at far distances $\bar{R}$, around the detector. The latter is described as a laser interferometer with bouncing photons $\gamma$. }
    \label{fig:detector}
\end{figure}

At this point, we adopt harmonic coordinates around the detector such that, in its vicinity, the Bondi coordinates can be expanded as
\begin{align}
\nonumber
X^i &= \bar{X}^i + x^i \,; \nonumber \\
U &= \bar{U} + u \,,
\label{local exp}
\end{align}
where the barred coordinates denote the center of mass and mean observation time of the detector, while small letters are perturbations about them, with~$u = t - \bar{N}_k x^k$. Without loss of generality, we can always set~$\bar{U} = 0$. See Fig.~\ref{fig:detector} for a schematic illustration. To leading order in~$1/R$, the GW strain of Eq.~\eqref{HijPN} then reads
\be
H_{ij}^\text{\tiny TT} (U,N) =   \frac{4G}{\bar{R}} \overline{\Pi}_{ij nr} \sum_{\ell = 2}^\infty \frac{1}{\ell!}  \bar{N}_{L-2}   \sum_{k = 0}^\infty  \frac{1}{k!}  \left( {\cal U}^{(k)}_{nr L-2}  + \bar{N}^s\epsilon_{ms(n}  {\cal V}^{(k)}_{r)mL-2} \right) u^k  \,,
\label{h rad}
\ee
where we have indicated time derivatives as~$f^{(k)} \equiv \frac{\partial^k}{\partial U^k} f(U) \vert_{U=\bar{U}}$. This expression describes a plane wave propagating in the~$\bar{N}^i$ direction. The first few terms are
\be
H_{ij}^\text{\tiny TT} (U,N) = \frac{1}{\bar{R}} \bigg( A_{ij}(\bar{N}) + B_{ij}(\bar{N}) u + \ldots\bigg)\,,
\label{hij TT local}
\ee
where~$A_{ij}(\bar{N})$ and~$B_{ij}(\bar{N})$ are both traceless and transverse to~$\bar{N}$. The reader can immediately appreciate that the time independent term~$A_{ij}$ can encode a displacement memory effect, induced by a change in the multipole moments $\mathcal{U}_{L}$ and $\mathcal{V}_{L}$.

The equivalence principle allows one to remove the leading terms displayed in Eq.~\eqref{hij TT local} by performing a suitable diffeomorphism~$\xi_i$,
with~$\delta H_{ij}^{\rm TT} = \partial_i\xi_j + \partial_j \xi_i$. To preserve TT gauge,~$\xi_i$ must satisfy~$\partial^i\xi_i = 0$,~$\vec{\nabla}^2 \xi_i = 0$. The necessary diffeomorphism satisfying these conditions is
\be
\xi_i = - \frac{1}{2\bar{R}}  \Big(A_{ij}  + B_{ij} t\Big) x^j   +  \frac{1}{4\bar{R}} \Big(B_{ij} \bar{N}_k + B_{ik} \bar{N}_j - B_{jk} \bar{N}_i\Big)x^j x^k\,.
\label{diff in terms of t}
\ee
The first piece, linear in~$\vec{x}$, describes a time-dependent volume-preserving (anisotropic) rescaling. The second piece, quadratic in~$\vec{x}$,
is the familiar Christoffel combination to remove a homogeneous acceleration. Notice that~$\xi_i$ can be equivalently written in terms of~$u$ as
\be
\xi_i = - \frac{1}{2\bar{R}}  \Big(A_{ij}  + B_{ij} u \Big) x^j - \frac{1}{4\bar{R}} B_{jk} \bar{N}_i x^j x^k\,.
\label{diff in terms of u}
\ee
Since~$\xi_i$ depends on time, we must perform a compensating time diffeomorphism~$\xi_0$ to preserve the TT gauge condition~$H_{0i}^{\rm TT}  = 0$. That is,~$\xi_0$ is chosen such that~$\delta H_{0i}^{\rm TT} = \partial_0\xi_i + \partial_i \xi_0 = 0$, with the solution  
\be
\xi_0 = \frac{1}{4\bar{R}} B_{ij} x^i x^j \,.
\label{time diff in terms of u}
\ee
This describes a spatially dependent time translation. 

For future use, the complete diffeomorphism can be written in the compact four-dimensional form
\begin{equation}
\label{diffa}
\xi^\mu = \sum_{n=1}^2 M^\mu_{\;\;  \mu_1 \cdots \mu_n} x^{\mu_1} \dots x^{\mu_n}\,, 
\end{equation}
where~$M^\mu_{\;\;  \mu_1 \cdots \mu_n}$ is symmetric in its last indices\footnote{Note that the transformation matrix~$M^\mu_{\;\;  \mu_1 \cdots \mu_n}$ should not be confused with the mass multipole moments~$M_L$.} and fully traceless, such that~$\partial_{\mu}\xi^{\mu}=0$ and~$\Box\xi^\mu = 0$.
Explicitly, from Eqs.~\eqref{diff in terms of t} and~\eqref{time diff in terms of u}, we have
\begin{eqnarray}
\nonumber
& n = 1:& ~~~~ M^i_{\,j}  = - \frac{1}{2\bar{R}} A^i_{\,j}\,; \\
\nonumber
& n = 2:& ~~~ M^i_{\,jk} = \frac{1}{4\bar{R}} \Big(B_{ij} \bar{N}_k + B_{ik} \bar{N}_j - B_{jk} \bar{N}_i\Big) \,; \\
& & ~~~ M^i_{\,j0} =  M^i_{\,0j} = - \frac{1}{4\bar{R}} B^i_{\, j}  \,; \qquad M^0_{\,ij} = - \frac{1}{4\bar{R}} B_{ij}  \,.
\label{M components explicit}
\end{eqnarray}
This induces the tensor mode~$\delta h_{\alpha \beta} = \partial_\alpha \xi_\beta + \partial_\beta \xi_\alpha$, given by
\begin{equation}
\label{modea}
\delta h_{\alpha \beta} = \sum_{n=1}^2  2n M_{\alpha \beta \mu_2 \dots \mu_n} x^{\mu_2} \dots x^{\mu_n}\,.
\end{equation}
This directly resembles Eq.~\eqref{hij TT local} and can  therefore be used to absorb the constant and linear contributions to the GW strain.

Let us stress that, because the local expansion of the GW strain in Eq.~\eqref{h rad} only depends on~$u$, one cannot remove any higher-order terms in the series expansion beyond the linear term. The quadratic term in~$u$ encodes the Riemann tensor, and it is therefore physical.\footnote{This can be seen from the geodesic deviation equation in TT gauge, $\ddot{\xi}_i = - R_{i0j0}\xi_j = \ddot{h}_{ij}^\text{\tiny TT} \xi_j/2$.} This is in contrast with the most general series expansion of~$h_{ij}$ in powers of~$x^\mu$,
where a subset of the coefficients at each order can be set to zero by a gauge-preserving diffeomorphism~\cite{Hinterbichler:2013dpa}.

\subsection{BMS transformation in TT coordinates}

Given that both BMS transformations and the local TT-preserving diffeomorphism in Eq.~\eqref{diffa} are able to describe the physics of the memory effects, it is worth showing their consistency. We do so here by properly expanding the BMS transformations in the local coordinate frame, and we show that the form of the corresponding tensor mode coincides with Eq.~\eqref{hij TT local}.

Our starting point is to cast the generalized BMS transformation in Eq.~\eqref{BMSdiffgeneral} in standard Minkowski coordinates.
Using the identities~\eqref{identities}, we obtain
\begin{align}
\xi_0^\text{\tiny BMS} &= \frac{R}{2} D_a Y^a - \frac{1}{2} \left(D^2 + 2\right) f\,;  \nonumber \\
\xi_i^\text{\tiny BMS} &=   \left( -\frac{R}{2} D_a Y^a + \frac{1}{2} D^2 f \right) N^i + e^i_a \Big(RY^a -D^a f\Big)\,,
\label{Mink diff}
\end{align}
where we recall that~$f = T + \frac{U}{2}  D_aY^a$. The super-Lorentz generator~$Y_a$ on the two-sphere admits a Helmhotz-Hodge decomposition in terms of two scalars~$\phi$ and~$\psi$ as
\be
Y_a = D_a \phi - \varepsilon_a^{\;c}D_c\psi = e^i_a R \Big(\partial_i \phi+ \epsilon_{ijk} N^j \partial_k \psi\Big)\,,
\ee 
where~$\varepsilon_{ac}$ is the Levi-Civita tensor on the sphere. Thus, the BMS transformation~\eqref{Mink diff} involves three scalar functions on the two-sphere:~$T$,~$\phi$, and~$\psi$.

It is useful to expand the above diffeomorphism in a multipolar STF decomposition: 
\begin{align}
\xi_0^\text{\tiny BMS} &= - \frac{R}{2} \sum_\ell \ell (\ell + 1) N_L \phi_L + \frac{1}{2} \sum_\ell (\ell + 2) (\ell -1) N_L\left(T_L - \frac{U}{2} \ell (\ell + 1)\phi_L\right)\,; \nonumber \\
\xi_i^\text{\tiny BMS} &= \frac{1}{2} N_i \sum_\ell \ell (\ell + 1) N_L \Big( R \phi_L  - T_L + \frac{U}{2} \ell (\ell + 1)\phi_L\Big) \nonumber  \\ 
&~~~ + P_{in} \sum_\ell \ell N_{L-1} \left( R\phi_{n L-1} + R N^k \epsilon_{nkm}  \psi_{m L-1} - T_{n L-1} +  \frac{U}{2} \ell (\ell + 1)\phi_{n L-1}\right)\,.
\label{BMS diff STF}
\end{align}
This diffeomorphism generates a long mode, with spatial components
\begin{align}
\nonumber
\delta h_{ij}^\text{\tiny BMS} &= 2  \Pi_{ijnr} \sum_\ell \ell (\ell -1) N_{L-2} \left[\phi_{nrL-2} + N^s\epsilon_{ms(n}  \psi_{r)m L-2}  + \frac{1}{R} \left(- T_{nrL-2} + \frac{U}{2} \ell (\ell + 1) \phi_{nrL-2}\right)  \right]  \\
\nonumber 
&+ \frac{1}{2} \Big(P_{in} N_j  + P_{jn} N_i \Big) \sum_\ell  \ell (\ell +2)(\ell-1) N_{L-1} \frac{1}{R} \left(- T_{nL-1} + \frac{U}{2} \ell (\ell + 1) \phi_{nL-1}\right) \\
&  -  \frac{1}{2} N_i N_j \sum_\ell \ell (\ell + 1) (\ell+2)(\ell -1) N_L \phi_L   \,,
\label{hij}
\end{align}
as well as time-space and time-time components given by
\begin{align}
\delta h_{0i}^\text{\tiny BMS} &= \frac{1}{2} N_i \sum_\ell \ell (\ell + 1)(\ell +2 ) (\ell -1) N_L \phi_L \nonumber \\
&+ \frac{1}{2} P_{in} \sum_\ell \ell (\ell + 2) (\ell -1) N_{L-1} \frac{1}{R}  \left(T_{nL-1} - \frac{U}{2} \ell (\ell + 1) \phi_{nL-1}\right)\,; \nonumber \\
\delta h_{00}^\text{\tiny BMS} & = -\frac{1}{2} \sum_\ell \ell (\ell + 1) (\ell + 2)(\ell -1) N_L \phi_L\,.
\label{h0i}
\end{align}
Since~$\delta h_{00}^\text{\tiny BMS}$ is time independent, it can be absorbed into a redefinition of the Newtonian potential, and we henceforth ignore it. 

In what follows, we will show that this long mode induced by a BMS transformation in TT coordinates is analogous to the one discussed in Eq.~\eqref{modea}. To make a straightforward comparison, let us expand the long mode of Eq.~$\eqref{hij}$ in the vicinity of the detector up to linear order in the coordinates:
\begin{align}
\delta h_{ij}^\text{\tiny BMS} = \hat{H}_{ij} + H_{ijk} x^k + Q_{ij} t + \ldots 
\label{hij expand}
\end{align}
The middle term can be simplified using the identity
\begin{equation}
H_{ijk} x^k = \bar{P}_{mk} H_{ijm}  x^k + \bar{N}^m H_{ijm}  \bar{N}_k x^k  = \bar{P}_{mk} H_{ijm}  x^k - Q_{ij}  \bar{N}_k x^k \,,
\end{equation}
where in the last step we have used the fact that~$\delta h_{ij}^\text{\tiny BMS}$ only depends on the combination~$u = t - \bar{N}_k x^k$; hence, we must have~$\bar{N}^m H_{ijm}  = -Q_{ij}$ automatically. 
Therefore, Eq.~\eqref{hij expand} becomes
\begin{equation}
\delta h_{ij}^\text{\tiny BMS} = \hat{H}_{ij} + \bar{P}_{km} H_{ijm} x^k + Q_{ij} u  + \ldots 
\label{hij expand final}
\end{equation}
Similarly, for~$\delta h_{0i}^\text{\tiny BMS}$ we have
\begin{equation}
\delta h_{0i}^\text{\tiny BMS} = \hat{H}_{0i} + \bar{P}_{mk} H_{0im} x^k + Q_{0i} u + \ldots 
\label{h0i expand}
\end{equation}
The above long mode induced by a BMS transformation in general takes us out of TT gauge. In order to restore TT gauge, we must perform a compensating diffeomorphism.
We will discuss the necessary coordinate transformations for the constant and linear-gradient terms, respectively, in the next two subsections.

\subsection{Constant mode}
\label{sec: constantmode}

Let us start from the constant mode in Eqs.~\eqref{hij expand final} and~\eqref{h0i expand}, with components~$\hat{H}_{ij}$ and~$\hat{H}_{0i}$.
From Eqs.~\eqref{hij} and~\eqref{h0i}, we explicitly have
\begin{align}
\nonumber
\hat{H}_{ij} &=  \sum_\ell \ell (\ell -1) \Bigg\{2  \overline{\Pi}_{ijnr}\bar{N}_{L-2} \left[\phi_{nrL-2} - \frac{1}{\bar{R}} T_{nrL-2} + \bar{N}^s\epsilon_{ms(n}  \psi_{r)m L-2}    \right]  \\
\nonumber 
& ~~~~~~~~~~~~~~~~~~~~     -\frac{1}{2\bar{R}} \Big(\bar{P}_{in} \bar{N}_j  + \bar{P}_{jn} \bar{N}_i \Big)  (\ell +2) \bar{N}_{L-1}  T_{nL-1}   -  \frac{1}{2} \bar{N}_i \bar{N}_j (\ell + 1) (\ell+2)\bar{N}_L \phi_L  \Bigg\} \,; \\
\hat{H}_{0i} &= \frac{1}{2}\sum_\ell \ell (\ell -1) (\ell +2 )\Bigg\{\bar{N}_i (\ell + 1)  \bar{N}_L \phi_L + \frac{1}{\bar{R} } \bar{P} _{in} \bar{N}_{L-1} T_{nL-1}\Bigg\} \,,
\label{Hij}
\end{align}
where we have set~$\bar{U} = 0$ without loss of generality. 

In order to restore TT gauge, we must perform a compensating diffeomorphism such that~$\hat{H}_{0i}$ vanishes,
while~$\hat{H}_{ij}$ becomes traceless and transverse to~$\bar{N}^j$. Starting with~$\hat{H}_{0i}$, the necessary compensating diffeomorphism
is a {\it time-dependent spatial translation} of the form
\be
\xi_i^\text{\tiny tran} = -\hat{H}_{0i} t \,.
\label{translation}
\ee
This readily implies~$\hat{H}_{0i}^\text{\tiny TT} = \hat{H}_{0i}  + \partial_0 \xi_i^\text{\tiny tran}  = 0$, as desired. 

Turning our attention to~$\hat{H}_{ij}$, notice that only the first line of Eq.~\eqref{Hij} is both traceless and transverse to~$\bar{N}^j$. The remainder of~$\hat{H}_{ij}$ can be removed by a {\it spatial rescaling}:
\be
\xi_i^\text{\tiny scaling} = \frac{1}{2} \left(\overline{\Pi}_{ijnr}\hat{H}_{nr} -\hat{H}_{ij}\right) x^j\,.
\label{spatial rescaling}
\ee
This leaves us with the first line of Eq.~\eqref{Hij} as the constant TT mode:
\be
\hat{H}_{ij}^\text{\tiny TT} = 2  \overline{\Pi}_{ijnr} \sum_\ell \ell (\ell -1) \bar{N}_{L-2} \left[\phi_{nrL-2} - \frac{1}{\bar{R}} T_{nrL-2} + \bar{N}^s\epsilon_{ms(n}  \psi_{r)m L-2}    \right]\,.
\ee
This takes the desired form of the constant term in Eq.~\eqref{h rad}, with the identification
\begin{align}
\phi_L - \frac{1}{\bar{R}}T_{L}   = \frac {2G}{\bar{R}\,\ell(\ell-1) \ell !}\, {\cal U}^{(0)}_{L}\,; \qquad 
\psi_{L} =  \frac{2G}{\bar{R} \ell \left(\ell -1\right)\ell!} \, {\cal V}^{(0)}_{L}  \,. 
\label{constant identification}
\end{align}
The explicit radial dependence in~$\phi_L$ and~$\psi_{L}$ shows that the necessary diffeomorphism depends on the location of the detector. Notice that~$\psi$ is fixed in terms of $\cal V$, in contrast with Eq.~\eqref{fUV}.

To summarize, at leading order in the coordinate expansion in the vicinity of the detector, a BMS diffeomorphism together with a compensating time-dependent translation~\eqref{translation} and spatial rescaling~\eqref{spatial rescaling},
\be
\xi_i = \xi_i^\text{\tiny BMS} + \xi_i^\text{\tiny tran} + \xi_i^\text{\tiny scaling} \,,
\ee
generates a constant TT mode of the desired form, {\it i.e.}, describing a constant shift induced by memory effects.  Furthermore, the reader can immediately recognize that the above coordinate transformation is compatible with the equivalence principle prediction of Eq.~\eqref{diffa}, $\xi_\mu = M_{\mu \mu_1} x^{\mu_1}$, thereby establishing the analogy between the diffeomorphisms in the two coordinate frames.

\subsection{Linear gradient mode}

Let us now discuss the linear gradient terms in Eq.~\eqref{hij expand final}, with coefficients~$H_{ijk}$ and~$Q_{ij}$.
From Eq.~\eqref{hij}, at linear order in~$x^i$, we have
\begin{align}
\label{Hijk BMS}
\nonumber
H_{ijk} &= \frac{1}{\bar{R}}\sum_\ell\ell\left(\ell -1\right)\Bigg\{  -\bar{N}_k   \ell \left(\ell + 1\right) \bigg[ \overline{\Pi}_{ijnr}   \bar{N}_{L-2} \phi_{nrL-2} + \frac{1}{4} \Big(\bar{P}_{jn}\bar{N}_i + \bar{P}_{in}\bar{N}_j \Big)  (\ell +2) \bar{N}_{L-1}  \phi_{nL-1}\bigg] \\
\nonumber
&- 2 \Big(\bar{N}_i \overline{\Pi}_{jknr} + \bar{N}_j \overline{\Pi}_{iknr}\Big) \bar{N}_{L-2} \Big(\phi_{nrL-2} + \bar{N}^s \epsilon_{ms(n}  \psi_{r)m L-2} \Big) \\
\nonumber
&-  \Big(\bar{P}_{in} \bar{P}_{jk} + \bar{P}_{jn} \bar{P}_{ik} - \bar{P}_{ij} \bar{P}_{kn} \Big)  \bar{N}_{L-1}  \Big(2 \phi_{nL-1} + \bar{N}^s \epsilon_{msn}  \psi_{m L-1} \Big) \\
\nonumber
&+ 2 \overline{\Pi}_{ijnr}\bigg[ \bar{P}_{kq}  (\ell -2) \bar{N}_{L-3} \Big(\phi_{nrqL-3} + \bar{N}^s \epsilon_{ms(n}  \psi_{r)mq L-3} \Big) + \bar{P}_{ks}  \bar{N}_{L-2} \epsilon_{ms(n} \psi_{r)m L-2}\bigg]\\
&-  \frac{1}{2}  (\ell + 1)(\ell + 2)  \bigg[    \Big(\bar{P}_{ik} \bar{N}_j  + \bar{P}_{jk} \bar{N}_i \Big)  \bar{N}_L \phi_L +  \bar{N}_i\bar{N}_j\bar{P}_{kn} \ell  \bar{N}_{L-1} \phi_{nL-1}\bigg]\Bigg\} \,.
\end{align}
Meanwhile, the coefficient of the term linear in~$t$, satisfying $Q_{ij} = - \bar{N}^k H_{ijk}$, is
\be
Q_{ij} =  \frac{1}{\bar{R}} \sum_\ell  \ell^2\left(\ell^2 -1\right) \Bigg[ \overline{\Pi}_{ijnr}   \bar{N}_{L-2} \phi_{nrL-2} + \frac{1}{4} \Big(\bar{P}_{jn}\bar{N}_i + \bar{P}_{in}\bar{N}_j \Big)  (\ell +2) \bar{N}_{L-1}  \phi_{nL-1}\Bigg] \,.
\label{Mij BMS}
\ee
In order to bring the linear-gradient mode into a form similar to Eq.~\eqref{h rad} and proportional to~$u$ only, all terms in Eq.~\eqref{Hijk BMS} except for the first line (proportional to~$\bar{N}_k$) must be canceled. This amounts to canceling the contributions~$\bar{P}_{k m}H_{ijm}$ in~\eqref{h0i expand}, which can be achieved with a compensating diffeomorphism describing a time-independent homogeneous acceleration:
\be
\xi_i^{(1)} = - \frac{1}{4} \Big( \bar{P}_{km} H_{ijm} +  \bar{P}_{jm} H_{ikm} - \bar{P}_{im} H_{jkm}\Big) x^j x^k \,. 
\label{comp 1}
\ee
As desired, this shifts the long mode by~$\delta H_{ijk} = -  \bar{P}_{km} H_{ijm}$ and cancels all but the first line. The linear gradient terms in Eq.~\eqref{hij expand final} then become
\be
\delta h_{ij}^\text{\tiny BMS}|_\text{\tiny lin} + 2 \partial_{(i}\xi^{(1)}_{j)} =  \frac{1}{\bar{R}} \sum_\ell  \ell^2\left(\ell^2 -1\right) \Bigg[ \overline{\Pi}_{ijnr}   \bar{N}_{L-2} \phi_{nrL-2} + \frac{1}{4} \Big(\bar{P}_{jn}\bar{N}_i + \bar{P}_{in}\bar{N}_j \Big)  (\ell +2) \bar{N}_{L-1}  \phi_{nL-1}\Bigg]  \, u\,.
\label{tilde h interm}
\ee
It remains to remove the term proportional to~$\bar{P}_{jn}\bar{N}_i + \bar{P}_{in}\bar{N}_j$, in order to be left with~$\delta h_{ij}$ proportional to the TT projector~$\overline{\Pi}_{ijnr}$, as in Eq.~\eqref{h rad}. This can be achieved with a second compensating diffeomorphism, given by
\be
\xi_i^{(2)} = \frac{x^j}{8\bar{R}} \Bigg[- \Big(\bar{P}_{jn}\bar{N}_i + \bar{P}_{in}\bar{N}_j\Big) t + \bar{P}_{in}\bar{N}_j \bar{N}_k x^k \Bigg]  \sum_\ell  \ell^2\left(\ell^2 -1\right)(\ell +2) \bar{N}_{L-1}  \phi_{nL-1} \,.
\label{comp 2}
\ee
This leaves us with the linear-gradient TT mode
\be
\label{hij+xi1}
\delta h_{ij}^\text{\tiny BMS}|_\text{\tiny lin} + 2 \partial_{(i}\xi^{(1)}_{j)} + 2 \partial_{(i}\xi^{(2)}_{j)} =  \frac{1}{\bar{R}} \overline{\Pi}_{ijnr}  \sum_{\ell=2}^\infty  \ell^2\left(\ell^2 -1\right) \bar{N}_{L-2} \phi_{nrL-2} \, u \,.
\ee
This matches the linear-gradient term of Eq.~\eqref{h rad}, provided that we fix
\be
\phi_{L}  =\frac{4G} {\ell^2 \left(\ell^2 -1\right)\ell!} \left({\cal U}^{(1)}_{L} +  \bar{N}^s  \epsilon_{ms(n}{\cal V}^{(1)}_{r)mL-2}\right) \,;\qquad \ell \geq 2\,.
\ee
Just like in Eq.~\eqref{constant identification} for the constant terms, this explicitly depends on the location of the detector because of the dependence on $\bar{N}$. Furthermore, notice that the linear-gradient multipole moments~${\cal U}^{(1)},{\cal V}^{(1)}$ completely fix the form of only the scalar~$\phi_L$, which is the only function entering at linear order in $U$ in the BMS diff of Eq.~\eqref{Mink diff}.

To fully restore TT gauge, it remains to eliminate~$\delta h_{0i}^\text{\tiny BMS}$, whose linear-gradient contribution can be read off from Eq.~\eqref{h0i}:
\begin{align}
H_{0ik} & = \frac{1}{2\bar{R}}\sum_\ell \ell \left(\ell -1\right)(\ell + 2)(\ell+1)  \Bigg\{\bar{P}_{ik}    \bar{N}_L \phi_L +  \frac{1}{2} \ell  \bar{N}_{L-1} \lp 2 \bar{N}_i \bar{P}_{nk} + \bar{N}_k \bar{P}_{in}  \rp \phi_{nL-1}\Bigg\} \nonumber \\
Q_{0i} & = - \bar{N}^k H_{0ik} = -  \frac{1}{4\bar{R}} \bar{P}_{in}  \sum_\ell \ell^2 \left(\ell -1\right) (\ell + 2)(\ell+1)  \bar{N}_{L-1}  \phi_{n L-1}\,.
\end{align}
Including an additional contribution from the time dependence of~$\xi_i^{(2)}$, we obtain
\begin{align}
\delta h_{0i}^\text{\tiny BMS}|_\text{\tiny lin} + 2\partial_{(0}\xi^{(2)}_{i)} & =  
\frac{1}{2\bar{R}}\sum_\ell \ell \left(\ell -1\right) (\ell + 2)(\ell +1) \Bigg\{   \bar{P}_{ik} x^k  \bar{N}_L   \phi_L   \nonumber \\
&~~~~ -  \frac{1}{2}\left[\bar{P}_{in} \left( u + \frac{1}{2} \bar{N}_k x^k\right) - \frac{3}{2} \bar{N}_i \bar{P}_{kn} x^k \right]  \ell  \bar{N}_{L-1}  \phi_{nL-1} \Bigg\}\,.
\label{h0i linear}
\end{align}
To set this quantity to zero, we must perform both a time and a spatial diffeomorphism, given by 
\begin{align}
\xi_0^{(3)} & =  - \frac{1}{4\bar{R}} x^i x^k \sum_\ell \ell \left(\ell-1\right) (\ell + 2)(\ell + 1) \Big(\bar{P}_{ik}\bar{N}_n + \ell  \bar{P}_{in}\bar{N}_k\Big) \bar{N}_{L-1}  \phi_{nL-1} \,;\nonumber \\
\xi_i^{(3)} & = \frac{1}{8\bar{R}}  \bigg[ \bar{P}_{in} t^2  + \Big( \bar{P}_{in} \bar{N}_k - \bar{P}_{kn} \bar{N}_i\Big) x^k t\bigg]  \sum_\ell  \ell^2\left(\ell -1\right)(\ell +2) (\ell + 1)\bar{N}_{L-1}  \phi_{nL-1} \,. 
\label{comp 3} 
\end{align}
The first term in~$\xi_i^{(3)}$, proportional to~$t^2$, describes a time-dependent spatial translation. The terms proportional to~$x^k t$, meanwhile, describe a time-dependent rotation. Being antisymmetric in~$i,j$, it does not contribute to~$\delta h_{ij}^\text{\tiny BMS}|_\text{\tiny lin}$ in Eq.~\eqref{hij+xi1}. Moreover, it is easy to see that the contributions from~$\xi_0^{(3)}$ and~$\xi_i^{(3)}$ to the~$(0,i)$ metric components precisely cancel out Eq.~\eqref{h0i linear}, as desired.

To summarize, the spatial diffeomorphism needed to absorb the linear term in Eq.~\eqref{h rad} is the sum of the BMS diffeomorphism together with the compensating transformations of Eqs.~\eqref{comp 1},~\eqref{comp 2}, and~\eqref{comp 3}. In other words, we have~$\xi_0^\text{\tiny lin} = \xi_0^\text{\tiny BMS} + \xi_0^\text{\tiny comp}$ and~$\xi_i^\text{\tiny lin} = \xi_i^\text{\tiny BMS} + \xi_i^\text{\tiny comp}$, with
\begin{align}
\xi_0^\text{\tiny comp} & = \xi_0^{(3)} =  - \frac{1}{4\bar{R}} x^i x^k \sum_\ell \ell \left(\ell -1\right) (\ell + 2)(\ell + 1) \Big(\bar{P}_{ik}\bar{N}_n + \ell  \bar{P}_{in}\bar{N}_k\Big) \bar{N}_{L-1}  \phi_{nL-1} \,; \nonumber \\
\xi_i^\text{\tiny comp} &= \xi_i^{(1)} + \xi_i^{(2)} + \xi_i^{(3)}  \nonumber \\
& = - \frac{1}{4} \Big( \bar{P}_{km} H_{ijm} +  \bar{P}_{jm} H_{ikm} - \bar{P}_{im} H_{jkm}\Big) x^j x^k \nonumber \\
& ~~~\, + \frac{1}{8\bar{R}} \Bigg[ \bar{P}_{in} u^2 + 2  \Big( \bar{P}_{in} \bar{N}_k - \bar{P}_{kn} \bar{N}_i\Big) x^k t \Bigg] \sum_\ell  \ell^2\left(\ell -1\right)(\ell +2)(\ell + 1) \bar{N}_{L-1}  \phi_{nL-1}  \,.
\end{align}
The first line in~$\xi_i^\text{\tiny comp}$ describes a time-independent homogeneous acceleration, familiar from the equivalence principle, while the second line is a combination of a~$u$-dependent spatial translation and a time-dependent rotation. Similarly to the constant piece, this is compatible with the prediction of Eq.~\eqref{diffa}, $\xi_\mu = M_{\mu \mu_1 \mu_2} x^{\mu_1} x^{\mu_2} $, showing the analogy between the diffeomorphisms in the two coordinate frames.

\section{Consistency relations for scattering amplitudes}

Having established the precise relation between asymptotic BMS symmetries and the large residual diffeomorphisms in the TT gauge familiar to cosmologists, we now 
show that gravitational soft theorems can be derived in the latter coordinates. Indeed, both the residual diffeomorphisms in cosmology and BMS transformations in asymptotically flat space-times give rise to soft theorems on their own~\cite{Creminelli:2012ed,Kehagias:2013yd,Peloso:2013zw, Hinterbichler:2013dpa, Hui:2018cag, Hui:2022dnm, Strominger:2013jfa, Creminelli:2024qpu}. 

In this Section, we derive the Ward-Takahashi identities for the residual diffeomorphisms in TT gauge given in given in Eq.~\eqref{diffa}, which we rewrite here for the convenience of the reader:
\begin{align} \label{eqn:diff}
   \xi^{\mu} = \sum_{n=1}^2 M^{\mu}_{\;\;  \mu_1 \cdots \mu_n} x^{\mu_1} \cdots x^{\mu_n}\,; \quad ~~~\text{with}~~~\partial_{\mu}\xi^{\mu}=0\,; \quad \Box\xi^\mu = 0\,.
\end{align}
We will show how these constrain the soft limits of scattering amplitudes with a soft graviton. In the process, we will see how Weinberg's soft factor~$\frac{1}{k \cdot q}$ arises from residual diffeomorphisms. To our knowledge, this had not been explicitly demonstrated before, without invoking polology arguments.

\subsection{Ward-Takahashi identity}
\label{WTI}

We focus on the field transformations due to the diffeomorphism $\xi = \xi^{\mu} \partial_{\mu} $, 
\begin{align}
\label{transf}
    \delta \psi & = {\cal L}_{\xi} \psi \,; \nonumber \\ 
    \delta h_{\mu\nu} & = \sum_{n=1}^2 2 n M_{(\mu\nu) \mu_2\dots \mu_n} x^{\mu_2} \cdots x^{\mu_n} +  {\cal L}_{\xi} h_{\mu\nu}\,,
\end{align}
where~$\psi$ represents a generic matter field.\footnote{Note that the matter field $\psi$ should not be confused with the scalar function~$\psi$ of the BMS diff.} Here,~${\cal L}_{\xi}$ is the Lie derivative with respect to~$\xi$, which for tensor modes takes the form~${\cal L}_{\xi} h_{\mu\nu} = \xi^\alpha \partial_\alpha h_{\mu\nu} + \partial_\mu \xi^\sigma h_{\sigma\nu} + \partial_\nu \xi^\sigma h_{\mu \sigma} $. To derive the Ward-Takahashi identity associated with this symmetry, consider the current~$Q^{\mu}=\xi_{\alpha} T^{\alpha \mu}$, where~$T^{\alpha \mu}$ contains  both the energy-momentum tensor of matter fields, as well as the pseudo energy-momentum tensor of~$h_{\mu\nu}$. The charge density~$Q^0$ satisfies the equal-time commutation relations
\begin{align}
   & {\rm i}\big[Q^0(t, \vec{x}), \psi(t, \vec{y}) \big] = \delta^{(3)}(\vec{x}-\vec{y}) {\cal L}_{\xi} \psi\,;   \nonumber \\
   & {\rm i}\big[Q^0(t, \vec{x}), h_{\mu\nu}(t, \vec{y}) \big] =\delta^{(3)}(\vec{x}-\vec{y}) {\cal L}_{\xi} h_{\mu\nu}\,.
\end{align}
In particular, we see that~$Q^0$ generates the linear part of~$\delta h_{\mu\nu}$, proportional to~${\cal L}_{\xi} h_{\mu\nu}$. In other words, $T^{\alpha 0}$ contains all the {\it hard charges}.\footnote{In contrast, soft charges only generate the nonlinear transformations of the field, similar to spontaneous symmetry breaking.} Introducing the field $\Phi$ to denote either~$\psi$ or~$h_{\mu\nu}$, one can write the identity
\begin{align}
\label{eqdcomm}
    {\rm i} \partial_{\mu} \vev{Q^{\mu}(x)  \Phi(x_1)\dots \Phi(x_N)} - & {\rm i} \vev{ \partial_{\mu} Q^{\mu}(x)  \Phi(x_1)\dots \Phi(x_N)}  \nonumber \\
   &\qquad  = \sum_{m=1}^N \delta^{(4)}(x-x_m) \vev{\Phi(x_1)\dots \delta\Phi(x_m)\dots  \Phi(x_N)}\,,
\end{align}
where~$\vev{\cdots}$ denotes time-ordered correlators built out of $N$ hard modes $\Phi$.\footnote{In deriving Eq.~\eqref{eqdcomm}, we have distributed the time derivative by following the standard procedure (see, {\it e.g.}, Chap.~10 of Ref.~\cite{Weinberg:1995mt}), based on considering the current associated with field transformations. The main difference is that in our case $\partial_\mu Q^{\mu} \neq 0$. The usual Ward-Takahashi identity with $\partial_\mu J^{\mu} = 0$ (upon using the equations of motion) reads, 
\begin{align} 
 {\rm i} \partial_{\mu} \vev{ J^{\mu}(x) \Phi(x_1)\dots \Phi(x_N) }=  \sum_{m=1}^N \delta^{(4)}(x-x_m) \vev{\Phi(x_1)\dots \delta\Phi(x_m)\dots  \Phi(x_N) }\,.
\end{align}}
Fourier-transforming both sides, using~$q^2=0$, gives
\begin{align} \label{eqn:WT}
\int {\rm d}^4x\, {\rm e}^{-{\rm i} qx }\bigg[ -q_{\mu} \vev{ Q^{\mu}(x) \Phi(x_1)\dots \Phi(x_N) } -& {\rm i}\vev{ \partial_{\mu} Q^{\mu}(x) \Phi(x_1)\dots \Phi(x_N) }\bigg] \nonumber \\
&\qquad = \sum_{m=1}^N {\rm e}^{ -{\rm i}qx_m} \vev{\Phi(x_1)\dots \delta\Phi(x_m)\dots  \Phi(x_N) }\,.
\end{align}
This provides the most general expression of the Ward-Takahashi identity. In the following, we will focus only on scalar fields as hard modes,~$\Phi = \varphi$, leaving the discussion of hard tensor modes to future work. With this assumption, the variation of the field is given by $\delta \varphi = \xi^{\mu} \partial_{\mu}\varphi$. Furthermore, our derivation of the subleading~($n=2$) soft theorem is strictly valid only at tree level. At the end of this Section we will comment about the generalization to loop corrections.

The first term on the left-hand side (lhs) of Eq.~\eqref{eqn:WT} can be neglected, if we keep only terms up to order~$\mathcal{O}(q)$ and realize that the correlator
does not have a pole at~$q = 0$, following~\cite{DiVecchia:2015jaq}. To deal with the second term on the lhs, we consider the Einstein equations
\begin{align}
\label{EOMGR}
   -2 \kappa T_{\mu \alpha}  = \Box h_{\mu \alpha} + \eta_{\mu \alpha} \partial^\gamma \partial^\delta h_{\gamma \delta} - \eta_{\mu \alpha} \Box h^\gamma_{\,\gamma} - \partial_{\mu} \partial^\gamma h_{\alpha \gamma } - \partial_{\alpha} \partial^\gamma h_{\mu \gamma } + \partial_{\mu} \partial_{\alpha} h^\gamma_{\,\gamma} \,,
\end{align}
 where $\kappa^2 = 8 \pi G$, and $h_{\mu\nu}$ has been canonically normalized. We apply the Schwinger-Dyson equation associated with these equations to replace~$-{\rm i}\vev{\partial_{\mu} Q^{\mu}(x)\dots }$ with $\frac{\rm i}{2 \kappa} \partial_{\mu}\xi_{\nu} \Box \vev{h^{\mu\nu}(x) \dots }$ in TT gauge. (Notice that only the first term on the right-hand side of Eq.~\eqref{EOMGR} is nonvanishing for this gauge choice.) The Fourier transform of the latter then simply becomes a Lehmann–Symanzik–Zimmermann (LSZ) operation~\cite{1955NCimS...1..205L} to obtain a graviton external state,\footnote{The form of the LSZ operator depends on the gauge choice for the graviton state. However, the result of the procedure, {\it i.e.}, the associated Ward identities for scattering amplitudes, is gauge invariant.}
\begin{align}
\frac{{\rm i}}{2\kappa}  \int {\rm d}^4x \, {\rm e}^{- {\rm i} qx } \partial_{\mu}\xi_{\nu} \Box \vev{h^{\mu\nu}(x) \dots } = - \frac{{\rm i}^{n}}{2\kappa} \sum_{n=1}^2 n M_{(\mu \nu) \mu_1 \cdots \mu_{n-1}} \frac{\partial}{\partial q^{\mu_1}} \cdots \frac{\partial}{\partial q^{\mu_{n-1}}} \vev{h^{\mu\nu}(q) | \dots }\,.
\end{align}
Notice that there are additional contact terms in the Schwinger-Dyson equation~\cite{Schwartz:2014sze}. However, these do not contribute in the scattering amplitudes after the LSZ reduction, as they contain fewer operators. Such contact terms vanish upon using momentum conservation, since all modes have nonzero momentum except for the soft graviton.

Next, we further perform a (deformed) LSZ reduction on both sides of Eq.~\eqref{eqn:WT} with all the insertions $x_l$,
\begin{align}
  \prod_{l=1}^N \lim_{\substack{k_l^2 \to - m_l^2\\ q\to 0}} \frac{{\rm i}\left((k_l+q)^2 + m_l^2 \right)}{\sqrt{Z_l}}  \int {\rm d}^4x \, {\rm e}^{-{\rm i} k_l x_l}\,.
\end{align}
For simplicity, we study only out-amplitudes~\cite{DiVecchia:2015jaq}. The difference between the usual LSZ and the deformed LSZ procedure is of~${\cal O}(q)$ after taking the limit, {\it i.e.}, up to terms of the form~$\frac{2 k_l \cdot q}{k_l^2 + m_l^2}$. Thus, the lhs of Eq.~\eqref{eqn:WT}, after applying the deformed LSZ procedure and up to terms of~${\cal O}(q)$, reduces to 
\begin{align} \label{eqn:LHS}
   \nonumber 
    \mbox{lhs of } \eqref{eqn:WT} &= - \sum_{n=1}^2 \lim_{q \to 0} \frac{n}{2 \kappa}{\rm i}^{N+n} M^{(\mu\nu) \mu_1 \dots \mu_{n-1}}  \\
&\;\;\;\;\;\;\;\;\times    \frac{\partial}{\partial q^{\mu_1}} \cdots \frac{\partial}{\partial q^{\mu_{n-1}}}   \bigg[ (2\pi)^4 \delta^{(4)} \Big(\sum\nolimits_m k_m +q \Big) {\cal T}_{\mu\nu}(q;k_1, \dots, k_N ) \bigg]\,,
\end{align}
where~${\cal T}_{\mu\nu}(q;k_1, \dots, k_N ) = \langle h_{\mu\nu}(q) \varphi(k_1)\cdots \varphi(k_N) |0 \rangle$ is the amplitude with a soft graviton~$h_{\mu\nu}(q)$. 

Similarly, the right-hand side (rhs) of Eq.~\eqref{eqn:WT}, after applying the deformed LSZ procedure and using Eq.~\eqref{transf}, becomes 
\begin{align} \label{eqn:RHS}
\mbox{rhs of } \eqref{eqn:WT} & = \sum_{n=1}^2 \sum_{l=1}^N  \lim_{\substack{k_l^2 \to - m_l^2\\ q\to 0}}  \big((k_l+q)^2 +m_l^2 \big) {\rm i}^{N+n} M^{\mu \mu_1 \cdots \mu_n} (k_l+ q)_{\mu} \nonumber \\  &~~~~~ \times\frac{\partial^n}{\partial k_l^{\mu_1} \cdots \partial k_l^{\mu_n}}\left[ \frac{(2\pi)^4 \delta^{(4)} \big(\sum_m k_m +q  \big) }{(k_l+q)^2 +m_l^2} {\cal T}(k_1, \dots, k_l+q,\dots, k_N)  \right]\,, 
\end{align} 
where ${\cal T}(k_1, \dots, k_l+q,\dots, k_N) = \langle \varphi(k_1)\cdots \varphi(k_l + q) \cdots \varphi(k_N)|0\rangle$ is the off-shell amplitude. In what follows, we are going to simplify both sides of the equation, considering the leading~$n = 1$ and subleading~$n =2$ terms of the general diffeomorphism in Eq.~\eqref{eqn:diff}.

\subsection{Leading soft theorem ($n = 1$)}

In the leading case~$n = 1$, the general diffeomorphism in Eq.~\eqref{eqn:diff} reduces to an anisotropic spatial rescaling,
\begin{align}
    \xi^{i} = M^{i}_{\; j} x^j\,,
\end{align}
where~$M_{ij}$ is symmetric and traceless, such that~$\partial_i\xi^i = 0$. From the~$n=1$ part of Eqs.~\eqref{M components explicit}, recall that~$M^i_{\;j}  = - \frac{1}{2\bar{R}} A^i_{\;j}$. In this case, the rhs of the Ward identity, given in  Eq.~\eqref{eqn:RHS}, reduces to the expression
\begin{align}
\mbox{rhs}  & = (2\pi)^4 \sum_l  \lim_{\substack{k_l^2 \to - m_l^2\\ q\to 0}} {\rm i}^{N+1} M^{ij} (k_l+ q)_{i} \nonumber \\
& ~~~~~~~~~~~ \left[    \Big((k_l+q)^2 +m_l^2 \Big)    \delta^{(4)} \Big(\sum\nolimits_m k_m +q \Big) \frac{\partial}{\partial k_l^j} \left( \frac{ {\cal T}(k_1, \dots, k_l+q,\dots, \bar{k}_N)   }{(k_l+q)^2 +m_l^2} \right)\right. \nonumber \\
 & ~~~~~~~~~~~ \left.+    {\cal T}(k_1, \dots, k_l+q,\dots, \bar{k}_N) \frac{\partial}{\partial k_l^j }  \delta^{(4)}\Big(\sum\nolimits_m k_m +q \Big)   \right]\,,
\label{rhs interm}
\end{align}
where we have used momentum conservation on the last leg to set
\be
\bar{k}_N = -\sum\limits_{l=1}^{N-1} k_l-q\,.
\label{bar kn}
\ee
The last line of Eq.~\eqref{rhs interm}, after performing an integration by parts for the derivative acting on the Dirac delta function, and after the on-shell limit and summation, vanishes up to ${\cal O}(q)$ terms, due to~$\partial_i \xi^i =0$ and $q+\sum_l k_l =0$.  Expanding out the middle line, and ignoring terms of~${\cal O}(q)$ that also arise from the expansion of~$\cal{T}$, we obtain 
\begin{align}
\mbox{rhs}  & = - (2\pi)^4 \delta^{(4)} \Big(\sum\nolimits_m k_m +q \Big) {\rm i}^{N+1} \nonumber \\
& ~~~~~~~\times \sum_{l=1}^N M^{ij} \left[  \frac{ (k_l+q)_i(k_l+q)_j}{k_l\cdot q} +  \frac{ k_{l\,i}k_{l\,j}}{k_l\cdot q}  q^{\mu}  \frac{\partial}{\partial k_{l}^{\mu} } - k_{l\,i} \frac{\partial }{\partial k_l^j} \right] {\cal T}(k_1, \dots, k_l,\dots, \bar{k}_N)\,.
\label{rhs n=1}
\end{align}
Meanwhile, the lhs of the identity, Eq.~\eqref{eqn:LHS}, with~$n=1$ gives
\begin{align}
\mbox{lhs} = - \frac{{\rm i}^{N+1}}{2\kappa}M^{ij} {\cal T}_{ij}(q; k_1, \dots, \bar{k}_N) (2\pi)^4 \delta^{(4)} \Big(\sum\nolimits_m k_m +q \Big) \,.  
\label{lhs n=1}
\end{align}
Combining Eqs.~\eqref{rhs n=1} and~\eqref{lhs n=1} gives the consistency relation for the leading, constant mode:
\begin{equation}
\label{eqn:idendityn=1}
    \boxed{
\begin{aligned} 
\frac{1}{2\kappa}M^{ij} {\cal T}_{ij}(q; k_1, \dots, \bar{k}_N)  =  \sum_{m=1}^N M^{ij} \left[  \frac{ (k_m+q)_i(k_m+q)_j}{k_m\cdot q} +  \frac{ k_{m\,i}k_{m\,j}}{k_m\cdot q}  q^{\mu}  \frac{\partial}{\partial k_{m}^{\mu} } - k_{m\,i} \frac{\partial }{\partial k_m^j} \right] {\cal T}(k_1, \dots, \bar{k}_N)\,.
\end{aligned}
}
\end{equation}
In this equality, we have removed the delta functions and implicitly assumed that each~$k_N$ should be replaced with Eq.~\eqref{bar kn}.
The leading pole in~$q$,~$\frac{1}{k_m\cdot q}$, is precisely Weinberg's soft factor for emitting/absorbing a soft graviton~\cite{Weinberg:1995mt}. It is important to emphasize that the appearance of the soft factor in the derivation above does not depend on using on-shell internal lines in Feynman diagrams. This also implies that the above statement should hold true nonperturbatively. Indeed, it was shown that the leading soft theorem does not receive loop corrections~\cite{Bern:2014oka}.

At $n=1$, one also finds additional linearly-realized symmetries, which are just boosts and rotations. They give rise to identities with only the rhs, of the form 
\begin{align}
\label{linearsoft}
   0=    \sum_m \bigg[ k_{m\,\mu} \frac{\partial }{\partial  k_{m}^{\nu}}  - k_{m\,\nu} \frac{\partial }{\partial k_{m}^{\mu}}  \bigg] {\cal T}(k_1, \dots, k_m,\dots, \bar{k}_N)\,.
\end{align}
We will make use of such identities to simplify the form of the subleading soft theorem below.

\subsection{Subleading soft theorem $(n = 2)$}

As discussed in Sec.~\ref{sec: diffTT}, the subleading term~$n = 2$ in the diffeomorphism of Eq.~\eqref{eqn:diff} takes the general form
\begin{align}
  \xi^{\mu} = M^{\mu}_{\;\; \alpha \beta} x^{\alpha} x^{\beta}\,,
\end{align}
where~$\mu,\alpha,\beta$ are space-time indices, and~$M^\mu_{\;\;  \alpha \beta}$ is symmetric in~$\alpha,\beta$ and fully traceless, which ensures that~$\partial_{\mu}\xi^{\mu}=0$ and~$\Box\xi^\mu = 0$. Its explicit components are given in Eqs.~\eqref{M components explicit}. 

With~$n=2$, the lhs of the Ward identity, given in  Eq.~\eqref{eqn:LHS}, becomes
\begin{align} \label{eqn:LHSn=2}
     {\rm lhs} 
    & =  \lim_{q \to 0} \frac{{\rm i}^{N}}{\kappa} M^{(\mu\nu) \alpha } (2\pi)^4 \left[ 
            \delta^{(4)} \Big(\sum\nolimits_m k_m +q \Big) \frac{\partial}{\partial q^{\alpha}} {\cal T}_{\mu\nu}(q;k_1, \dots, k_N ) \right. \nonumber \\ 
      & ~~~~~~~~~~~~~~~~~~~~~~~~~~~~~~~~~ + \left.\frac{\partial}{\partial q^{\alpha}} \delta^{(4)} \Big(\sum\nolimits_m k_m +q \Big) {\cal T}_{\mu\nu}(q;k_1, \dots, k_N ) \right]\,.
\end{align}
The rhs of the Ward identity in Eq.~\eqref{eqn:RHS} instead reduces to the more involved expression
\begin{align} 
\label{eqn:n=2full}
{\rm rhs}  & = - {\rm i}^{N} \sum_{l=1}^N  \lim_{\substack{k_l^2 \to - m_l^2\\ q\to 0}} M^{\mu\alpha\beta } (k_l+ q)_{\mu}  \Big((k_l+q)^2 +m_l^2 \Big)    (2\pi)^4 \nonumber \\
 & ~~~~~~~~~~~~~~~\times \left[    \delta^{(4)} \Big(\sum\nolimits_m k_m +q \Big) \frac{\partial^2 }{\partial k_l^\alpha \partial k_l^\beta } \left( \frac{ {\cal T}_{N,l}   }{(k_l+q)^2 +m_l^2} \right) \right. \nonumber \\
 &~~~~~~~~~~~~~~~~~~~~+2   \frac{\partial }{\partial k_l^\alpha } \left( \frac{ {\cal T}_{N,l} }{(k_l+q)^2 +m_l^2} \right)  \frac{\partial }{\partial k_l^\beta } \delta^{(4)} \Big(\sum\nolimits_m k_m +q \Big)
 \nonumber \\
& ~~~~~~~~~~~~~~~~~~~~ \left. +  \frac{ {\cal T}_{N,l}  }{(k_l+q)^2 +m_l^2}\frac{\partial^2 }{\partial k_l^\alpha \partial k_l^\beta } \delta^{(4)} \Big(\sum\nolimits_m k_m +q \Big) \right]\,,
\end{align} 
where we have abbreviated the off-shell scattering amplitude as ${\cal T}_{N,l}={\cal T}(k_1, \dots, k_l+q,\dots, k_N) $. Among these three terms, we will prove that only the first one contributes to the subleading relation.  

To see this, we first show that the last line does not contribute up to ${\cal O}(q)$. Since delta functions are distributions,
we can introduce the generalized momentum $P_{\mu} = \big(\sum_m k_m +q\big)_{\mu}$ and a test function $F(\{k_m\})$, such that the last line of~\eqref{eqn:n=2full}, together with factors outside the square bracket, reads 
\begin{align}
& (2\pi)^4  \sum_l   M^{\mu\alpha\beta } (k_l+ q)_{\mu}  {\cal T}_{N,l} \frac{\partial^2 }{\partial k_l^\alpha \partial k_l^\beta } \delta^{(4)} \Big(\sum\nolimits_m k_m +q \Big)  \nonumber  \\
& =   (2\pi)^4   M^{\mu\alpha\beta } \left. \Bigg( \sum_l  (k_l+ q)_{\mu} {\cal T}_{N,l}\Bigg) \right|_{\sum_m k_m +q=0}  \frac{\partial^2 }{\partial P_{\alpha} \partial P_{\beta} } \delta^{(4)} \Big(\sum\nolimits_m k_m +q \Big)  \nonumber \\
&  -(2\pi)^4  2 M^{\mu\alpha\beta } \frac{\partial }{\partial P_{\alpha} }  \left. \bigg( \sum_l (k_l+ q)_{\mu} {\cal T}_{N,l}\bigg) \right|_{\sum_m k_m +q=0}  \frac{\partial }{\partial P_{\beta} } \delta^{(4)} \Big(\sum\nolimits_m k_m +q \Big) \nonumber \\
&  +(2\pi)^4  M^{\mu\alpha\beta } \frac{\partial^2 }{\partial P_{\alpha} \partial P_{\beta}  } \left. \bigg( \sum_l (k_l+ q)_{\mu} {\cal T}_{N,l}\bigg) \right|_{\sum_m k_m +q=0}   \delta^{(4)} \left(\sum\nolimits_m k_m +q \right)\,,
\label{Eq.84}
\end{align}
where an implicit on-shell ($k_l^2 \to - m_l^2$) and soft limit ($q \to 0$) has been assumed.
The above expression vanishes up to ${\cal O}(q) $, thanks to momentum conservation~$\sum_l k_l +q=0$ or using the property of the transformation matrix of being traceless, $M^{\mu\alpha}_{\quad \alpha}=0$. Notice that terms proportional to $\partial F/\partial P_\alpha$  are simplified as well because of these two properties, and that we have ultimately removed the test function from both sides of Eq.~\eqref{Eq.84}.

Let us now study the second line of Eq.~\eqref{eqn:n=2full}, proportional to~$\frac{\partial }{\partial k_l^\beta } \delta^{(4)} \big(\sum_m k_m +q \big)$. By combining it with the second term on the lhs from Eq.~\eqref{eqn:LHSn=2}, one gets
\begin{align} 
\frac{1}{2\kappa} M^{\mu \alpha \beta} {\cal T}_{\mu\alpha}(q;k_1, \dots, k_N )  + \sum_{l=1}^N  \lim_{\substack{k_l^2 \to - m_l^2}} M^{\mu\alpha\beta } (k_l+ q)_{\mu}  \Big((k_l+q)^2 +m_l^2 \Big)    \frac{\partial }{\partial k_l^\alpha } \left( \frac{ {\cal T}_{N,l} }{(k_l+q)^2 +m_l^2} \right)\,,
\end{align}
where we have factored out the term $(2 \pi)^4\frac{\partial }{\partial P^\beta } \delta^{(4)} \big(\sum_m k_m +q \big)$ and omitted the small $q$ limit, for simplicity. By expanding the sum over the indices $\mu, \alpha$, and noticing that $M^{00\alpha} = 0$ and $M^{0}_{i\alpha} = - M^{i}_{0\alpha}$, in TT gauge the expression simplifies to
\begin{align} 
\frac{1}{2\kappa} M^{i j \beta} {\cal T}_{i j}(q;k_1, \dots, k_N )   & +  \sum_{l=1}^N  \lim_{\substack{k_l^2 \to - m_l^2}} \Bigg\{ 
 M^{i j\beta } (k_l+ q)_{i}  \Big((k_l+q)^2 +m_l^2 \Big)    \frac{\partial }{\partial k_l^j } \left( \frac{ {\cal T}_{N,l} }{(k_l+q)^2 +m_l^2} \right)  \nonumber \\
&~~~~~~~~~~~~~~~~~~~  + M^{0}_{i \beta}  \left[ (k_l+ q)_{0}  \Big((k_l+q)^2 +m_l^2 \Big)    \frac{\partial }{\partial k_l^i } \left( \frac{ {\cal T}_{N,l} }{(k_l+q)^2 +m_l^2} \right) \right. \nonumber \\
&~~~~~~~~~~~~~~~~~~~
\left.  - (k_l+ q)_{i}  \Big((k_l+q)^2 +m_l^2 \Big)    \frac{\partial }{\partial k_l^0 } \left( \frac{ {\cal T}_{N,l} }{(k_l+q)^2 +m_l^2} \right) \right]  \Bigg\} \,.
\end{align}
By including the linear-realized symmetries of Eq.~\eqref{linearsoft}, it is easy to show that the last two lines vanish. We thus reproduce an expression that is proportional to the leading soft theorem of Eq.~\eqref{eqn:idendityn=1}, which therefore vanishes as well.

We are therefore left with the first lines of Eqs.~\eqref{eqn:LHSn=2} and~\eqref{eqn:n=2full}.
The latter can be simplified using the identity
\begin{align}
& \frac{\partial^2 }{\partial k_l^\alpha \partial k_l^\beta } \left( \frac{ {\cal T}_{N,l}   }{(k_l+q)^2 +m_l^2} \right) = \frac{1}{(k_l+q)^2 + m_l^2} \Bigg[\frac{\partial^2 }{\partial k_l^\alpha \partial k_l^\beta } - \frac{2 (k_l + q)_\alpha}{(k_l+q)^2 + m_l^2} \frac{\partial}{\partial k_l^\beta} - \frac{2 (k_l + q)_\beta}{(k_l+q)^2 + m_l^2} \frac{\partial}{\partial k_l^\alpha} \nonumber \\
& \left.~~~~~~~~~~~~~~~~~~~~~~~~~~~~~~~~~~~~~~~~~~~~~~~~~~  + \frac{8 (k_l + q)_\beta (k_l + q)_\alpha -  2 \delta_{\alpha\beta} \left((k_l + q)^2 + m_l^2\right)}{\left[(k_l+q)^2 + m_l^2\right]^2} 
\right] {\cal T}_{N,l}\,.
\end{align}
Further expanding the expression up to~${\cal O}(q)$, we obtain
\begin{align}
 {\rm rhs} & = - \lim_{q \to 0} {\rm i}^{N} (2\pi)^4 \delta^{(4)} \Big(\sum\nolimits_m k_m +q \Big)\sum_{l=1}^N   M^{\mu\alpha\beta }  (k_l+q)_\mu  \nonumber \\
 &~~~~~~~~~~  \times\left[ 2 \frac{(k_l+q)_{\alpha}(k_l+q)_{\beta}}{( k_l \cdot q)^2}\left( 1 + q^{\nu} \frac{\partial}{\partial k_l^\nu} \right)  + \frac{\partial^2 }{\partial k_l^\alpha \partial k_l^\beta }  - 2\frac{(k_l + q)_\alpha}{k_l\cdot q} \frac{\partial}{\partial k_l^\beta} \right] {\cal T}_N\,.
\end{align}
Equating this to the first line of Eq.~\eqref{eqn:LHSn=2} gives us the subleading consistency relation:
\begin{equation}
\label{eqn:idendityn=2}
    \boxed{
\begin{aligned}
&\frac{1}{\kappa}  M^{(\mu\nu) \alpha } \frac{\partial}{\partial q^{\alpha}} {\cal T}_{\mu\nu}(q;k_1, \dots, \bar{k}_N)  \\
&~~~~~~~~= -  \sum_{m=1}^N  M^{\mu\alpha\beta } (k_m+q)_\mu  \left[ 2 \frac{(k_m+q)_{\alpha}(k_m+q)_{\beta}}{( k_m \cdot q)^2}\left( 1 + q^{\nu} \frac{\partial}{\partial k_m^\nu} \right)  \right.  \\
 & \hspace{6cm} \left. + \frac{\partial^2 }{\partial k_m^\alpha \partial k_m^\beta } - 2\frac{(k_m + q)_\alpha}{k_m\cdot q} \frac{\partial}{\partial k_m^\beta}   \right] {\cal T}(k_1, \dots, \bar{k}_N)\,.
\end{aligned}
}
\end{equation}
As with the leading consistency relation in Eq.~\eqref{eqn:idendityn=1}, we have removed the delta functions, and implicitly assumed that each~$k_N$ should be replaced with Eq.~\eqref{bar kn}. It is worth emphasizing again that the soft factor $\frac{1}{ k_j \cdot q}$ was obtained without the use of nearly on-shell internal propagators in diagrammatic arguments. 

Before closing this Section, it is instructive to compare the above result with that of Ref.~\cite{Hamada:2018vrw}, which shares the same symmetries for the graviton sector. The form of the Ward-Takahashi identity in their work is similar to our Eqs.~\eqref{eqn:LHS} and~\eqref{eqn:RHS}, even though their expression keeps the propagators and vertices written in compact form. In contrast, in this work we have expanded the expression and arrived at an explicit form containing all relevant soft factors consistently up to~${\cal O}(q)$. 

Finally, let us comment about the validity of these results beyond the tree-level assumption. Several works have shown that, contrarily to the leading soft theorem, the subleading relation receives corrections at one loop induced by the long-range nature of gravitational interactions~\cite{Laddha:2018myi, Sahoo:2018lxl, Laddha:2018vbn,  Campiglia:2019wxe, Saha:2019tub, Krishna:2023fxg}. Such nonanalytic corrections, proportional to $\log \omega$ (with $\omega = q^0$ being the frequency of the associated soft mode), may become more relevant than the $\mathcal{O}(q)$ terms of the tree-level subleading theorem and cast a veil of ambiguity on the subleading infrared triangle. It was later understood that these infrared divergent pieces were universal, and can be related to tails-of-memory effects  for massive fields in classical gravity~\cite{Laddha:2018vbn,Ghosh:2021bam,Sen:2024bax, Geiller:2024ryw}. These effects appear in the gravitational shear tensor as $H_{ij}^\text{\tiny TT} \supset C_{ij}/u$ in Eq.~\eqref{h rad} and  consist of a combination of tail effects (describing the backscattering of linear GWs against the curvature of space-time generated by the source) and the memory effect. As such, they are sourced by mass-quadrupole-quadrupole couplings and enter the waveform at the 4PN order~\cite{Trestini:2022tot, Trestini:2023ssa}. The completion of a subleading infrared triangle, including the logarithmic soft theorem and tails-of-memory effects, was recently accomplished through the  superrotation symmetry~\cite{Donnay:2022hkf, Agrawal:2023zea, Choi:2024ygx, Choi:2024ajz}, in complete analogy to supertranslations, displacement memory and leading soft theorems.
These papers showed that the Ward identity of superrotation symmetries, associated with the  conservation law of their charge across spatial infinity, was able to reproduce the classical logarithmic soft graviton theorem upon considering a dressing of the massive hard modes induced by their long-range gravitational interactions with the soft graviton~\cite{Giddings:2019hjc}.

In our computation, the residual diff of Eq.~\eqref{eqn:diff} properly captures the effects of superrotations (see, {\it e.g.}, Eq.~\eqref{Mink diff}), which are responsible for the tree-level subleading soft theorem in~\eqref{eqn:idendityn=2}. However, in the derivation of Sec.~\ref{WTI}, the current $Q^\mu = \xi_\alpha T^{\alpha \mu}$ is built only considering the energy-momentum tensor of free massive hard fields, which induces a linear variation in the right-hand side of the soft theorems. In order to consider loop corrections, associated with the logarithmic divergence $\log \omega$, one would need to perform a dressing of the free fields along the lines of Refs.~\cite{Donnay:2022hkf, Agrawal:2023zea, Choi:2024ygx, Choi:2024ajz} to properly track the long-range nature of the gravitational interactions with the soft graviton. We plan to investigate this direction in future work.

\section{Consistency relations for correlators}

Consistency relations are exact symmetry statements in cosmology. Analogously to the known soft theorems in high energy physics, they relate an~$(N+1)$-point correlation function in the ``squeezed" limit to an~$N$-point function. They are usually associated with the existence of nonlinearly realized symmetries in the theory, providing deep information on its content. Consistency relations have been studied in various cosmological contexts, such as inflation~\cite{Maldacena:2002vr, Creminelli:2004yq, Cheung:2007sv} and large scale structure~\cite{Kehagias:2013yd,Peloso:2013zw}, and hold for both scalar and tensor soft modes~\cite{Maldacena:2002vr,Creminelli:2004yq,Cheung:2007sv,Creminelli:2012ed,Hinterbichler:2013dpa}. They have also been derived for asymptotic symmetries of cosmological space-times~\cite{Mirbabayi:2016xvc, Ferreira:2016hee, Hinterbichler:2016pzn}. 

It is natural to show the existence of consistency relations also in the context of the large residual diffeomorphisms in TT gauge, found in Sec.~\ref{sec: diffTT}. Various techniques have been used to derive cosmological consistency relations, including the background wave method~\cite{Maldacena:2002vr, Creminelli:2004yq}, Ward identities~\cite{Hinterbichler:2013dpa}, the effective action~\cite{Goldberger:2013rsa}, the wave functional~\cite{Pimentel:2013gza, Kundu:2015xta}, Slavnov-Taylor identities~\cite{Berezhiani:2013ewa,Binosi:2015obq}, and the path integral approach~\cite{Hui:2018cag}. We will follow the latter approach, based on~\cite{Hui:2018cag}, summarized in the Appendix. In this regard, a related work is Ref.~\cite{Creminelli:2024qpu}, which considered gravitational memory and TT gauge residual diffeomorphisms for primordial tensor perturbations.

Consider a general equal-time operator~${\cal O}(\vec{k}_1,\dots, \vec{k}_N)$ built out of~$N$ generic fields~$\Phi$. The consistency relation connecting the~$N+1$ in-in correlator with an external soft mode to the $N$-point function, based on some nonlinearly realized symmetry, is shown in Eq.~\eqref{eqn:eqTidentityQ} to be of the general form
\begin{align}
     {\cal D}_{\vec q} \left. \frac{\vev{ {\cal O}(\vec{k}_1,\dots, \vec{k}_N)   \Phi \left(\vec{q} \right) }_{\rm c}^\prime }{\vev{ \Phi \left(\vec{q} \right) \Phi \left(-\vec{q} \right)}' }\right|_{\vec q =0} = \vev{\delta_\text{\tiny L}{\cal O}(\vec{k}_1,\dots, \vec{k}_N) }_{\rm c}^\prime\,,
\label{gen identity}
\end{align}
where the subscript ``c" denotes the connected part of correlators, introduced to remove terms associated with~$\delta_\text{\tiny NL}{\cal O}$ on the rhs of Eq.~\eqref{gen identity}, as shown in Appendix~C of Ref.~\cite{Hinterbichler:2013dpa}. Primes indicate the removal of three-dimensional Dirac delta functions enforcing momentum conservation.
The operator~${\cal D}_{\vec q}$ is associated with the nonlinear variation of the fields, as displayed in Eq.~\eqref{Dq}, while~$\delta_\text{\tiny L}$ denotes their linear variation. 

The large residual diffeomorphisms in TT gauge associated with GW memory are given by Eq.~\eqref{diffa}. In the following, we will focus first on the leading-order ($n=1$) contribution, which describes an anisotropic spatial rescaling:
\begin{align}
    \xi^i = M^{i}_{\,j} x^j \,,
\label{aniso diff}
\end{align}
where~$M_{ij}$ is symmetric and traceless. This can remove the long wavelength limit of a gravitational wave~$h_{ij}$ in TT gauge,~$ h_{ij} \to \bar{h}_{ij} = 2M_{ij}$, which mimics a constant memory term. Focusing on the time-independent part of the long mode, the equal-time correlation of~${\cal O}$  with the soft mode should be equivalent to the correlation of~${\cal O}$ evaluated in the transformed coordinate~$\tilde{x}^{i} =x^{i}+ \xi^{i}(x)$:
\begin{align}
\vev{{\cal O}(x_1,\ldots,x_N)}_{h \to {\rm const.}} & = \vev{{\cal O}(\tilde{x}_1,\ldots,\tilde{x}_N)} \,,
\label{Npt identity}
\end{align}
where~$\langle \ldots \rangle_{h}$ is the correlator with the long mode, and~$\langle \ldots \rangle$ is the correlator without.

Let us focus for concreteness on an observable~${\cal O}(x_1,\dots, x_N)$ built from a spectator scalar field~$\varphi$, which under~\eqref{aniso diff} transforms as $\delta \varphi =  M^{i}_{\;\;j} x^j \partial_i \varphi$. In the presence of a TT gravitational plane wave~$h_{ij}$, the right-hand side of Eq.~\eqref{Npt identity} can be expanded with respect to~$M_{ij} = \frac{1}{2}\bar{h}_{ij}$ as
\begin{align}
    \vev{\mathcal{O}(\tilde{x}_1,\ldots \tilde{x}_N)}  =  \left(1 + \frac{1}{2}\bar{h}^k_{\;\ell} \sum_{m = 1}^N x_m^\ell \frac{\partial}{\partial x_m^k} + {\cal O}\big(\bar{h}^2\big) \right)\vev{{\cal O}(x_1,\ldots,x_N)}\,, 
\end{align}
Correlating both sides with~$h_{ij}$ gives\footnote{To correlate with~$h_{}$, one can use the following relation
\begin{equation}
\langle h \langle \mathcal{O} \rangle_h \rangle = \int {\rm d} h \, h \int {\rm d} \mathcal{O} \, \mathcal{O} P(\mathcal{O}|h) P(h) = \int {\rm d} h \, {\rm d} \mathcal{O} \, h \mathcal{O} P(h,\mathcal{O}) = \langle h  \mathcal{O} \rangle \,,
\end{equation}
in terms of the probability distribution $P$ of the fields.}
\begin{align}
\label{CRcorrelators}
 \lim_{h \to {\rm const.}} \vev{h_{ij}{\cal O}(x_1,\ldots,x_N)} =   \lim_{h \to {\rm const.}} \frac{1}{2}\vev{h_{ij} \bar{h}_{k\ell}} \sum_{m = 1}^N x_m^\ell \frac{\partial}{\partial x_m^k} \vev{{\cal O}(x_1,\ldots,x_N)}\,,
\end{align}
 where we have distinguished the field~$h$ from the mode~$\bar{h}$ being removed by the diffeomorphism. Equation~\eqref{CRcorrelators} represents the consistency relation for scalar correlators under the anisotropic spatial rescaling symmetry in synchronous coordinates.

It should be stressed that, in the last step, it is not always guaranteed that the field $h$ in the soft limit is correlated with the constant profile $\bar{h}$. A necessary condition is the physical mode condition~\cite{Hui:2018cag}, according to which the time dependence of $h$ in the soft limit must match the time dependence of $\bar{h}$. For instance, both the slow-roll and ultra-slow-roll models  have a dilation symmetry~\cite{Finelli:2017fml}, but only the former satisfies the equal-time dilation consistency relation. In an ultra-slow-roll model, the soft mode $\zeta$ does not match the time dependence of the dilation generated constant mode $\bar{\zeta}$.

One further constraint must be imposed on these transformations to describe physically realized symmetries. In order for the shift in~$h_{ij}$, induced by the diffeomorphism~\eqref{aniso diff} as shown in Eq.~\eqref{transf}, to correspond to the long-wavelength of a physical mode,~$M_{ij}$ should satisfy the transversality condition in momentum space:
\be
\hat{q}^i M_{ij} (\hat{q}) = 0\,.
\ee
This condition is the analog of the ``adiabatic" mode condition in cosmology~\cite{Hinterbichler:2013dpa}. It reduces the number of independent components of~$M$ from five to two, which matches the number of physically propagating modes. With this fact at hand, the identity in Eq.~\eqref{gen identity} for the complex operator~${\cal O}(\vec{k}_1,\dots, \vec{k}_N) = \varphi(\vec{k}_1)\cdots \varphi(\vec{k}_N)$ reads explicitly 
\begin{align}
    \lim_{\vec q \to 0} \Pi^{ijk\ell}(\hat{q}) \frac{1}{P_h(q)} \left\langle h_{k\ell}(\vec q)  \varphi(\vec{k}_1)\cdots \varphi(\vec{k}_N)\right\rangle_{\rm c}' = -  \Pi^{ijk}_{~~\,\, \ell}(\hat{q})\sum_{m = 1}^N k_m^\ell \frac{\partial}{\partial k_m^k} \left\langle \varphi(\vec{k}_1)\cdots \varphi(\vec{k}_N)\right\rangle_{\rm c}'\,.
\label{N=2 identity with scalars}
\end{align}
We removed the $M_{ij}(\hat{q})$ coefficients by projecting the contracted indices on the TT subspace using the TT projector~$\Pi^{ijmn}(\hat{q})$. (This projector is given by
Eq.~\eqref{TT projector}, with~$P_{ij}(\hat{q}) = \delta_{ij}- \hat{q}_i \hat{q}_j$.) The quantity $P_h(q)$ denotes the tensor power spectrum, defined as the two-point correlation function of the Fourier-transformed tensor perturbations $\vev{h_{ij} (\vec{q}) h_{kl} (\vec{q}\,')} = (2\pi)^3 \delta^{(3)} (\vec{q} + \vec{q}\,')  2 \Pi_{ijkl} (\hat{q}) P_h(q)$. 

Similar relations can be deduced for correlators comprised of hard tensor modes, although the structure of their linear variation is more complex, owing to the requirement of preserving TT gauge. See Refs.~\cite{Hinterbichler:2013dpa, Berezhiani:2014tda} for further details.

At this point, we can generalize the soft theorems for cosmological correlations by including the contribution from the subleading linear gradient diff, which reads
\begin{equation}
 \xi^\mu = M^\mu_{\,\,\, \nu \rho} x^\nu x^\rho\,,  
\end{equation}
where $M^\mu_{\,\,\, \nu \rho}$ is  symmetric in its last indices and fully traceless.

To do so, we will follow similar steps to the ones considered for the leading term, and extend the relation~\eqref{Npt identity}, between the equal-time correlation of $\mathcal{O}$ with the soft mode and the correlation of $\mathcal{O}$ in the transformed coordinates $\tilde{x}^\mu = x^\mu + \xi^\mu (x) = x^\mu + M^\mu_{\,\,\, \nu \rho} x^\nu x^\rho$, to
\begin{align}
\label{OOsub}
\hspace{-0.3cm}\left\langle {\cal O}(x_1,\ldots,x_N)\right\rangle_{h \to \frac{1}{\bar{R}}(A + B u)} = \left\langle{\cal O}(\tilde{x}_1,\ldots,\tilde{x}_N)\right\rangle\,.
\end{align}
In this relation, we wish to take into account the linear gradient contribution of the long-wavelength part of a gravitational
wave $h_{ij}$ in TT gauge, $h_{ij} \to \frac{1}{\bar{R}} (A_{ij} + B_{ij} u)$ (see Eq.~\eqref{hij TT local}). Then, the right-hand side of~\eqref{OOsub} reads
\begin{align}
    \left\langle \mathcal{O}(\tilde{x}_1,\ldots, \tilde{x}_N)\right\rangle  = \left(1 + M^\mu_{\,\,\, \nu \rho} \sum_{m = 1}^N x_m^\nu x_m^\rho \frac{\partial}{\partial x_m^\mu} \right)\left\langle{\cal O}(x_1\ldots,x_N)\right\rangle\,, 
\end{align}
where we have considered the linear variation~$\delta \varphi = M^\mu_{\,\,\, \nu \rho} x^\nu x^\rho \partial_\mu \varphi$ of the scalar fields building the equal-time operator~$\mathcal{O}$. Correlating both sides of Eq.~\eqref{OOsub} with~$h_{ij}$ then gives
\begin{align}
\label{CRcorrelators-sub}
 \lim_{h \to \frac{1}{\bar{R}}(A + B u)} \left\langle h_{ij}{\cal O}(x_1,\ldots,x_N) \right\rangle  =   \lim_{h \to \frac{1}{\bar{R}}(A + B u)}  \left\langle h_{ij} M^\mu_{\,\,\, \nu \rho} \right\rangle \sum_{m = 1}^N x_m^\nu x_m^\rho \frac{\partial}{\partial x_m^\mu} \left\langle{\cal O}(x_1,\ldots,x_N)\right\rangle\,.
\end{align}
Equation~\eqref{CRcorrelators-sub} represents the consistency relations for correlators under the subleading memory-induced symmetry in TT coordinates.

In analogy to the physical mode condition discussed below Eq.~\eqref{CRcorrelators}, even at the subleading order one has to assume that the time dependence of $h$ in the soft limit matches the time dependence of the mode~$M x \sim B x$, to ensure that their correlation is nonvanishing.\footnote{This condition is not always guaranteed: for example, in models of relativistic superfluids, the subleading identities associated with boost symmetries cannot be promoted to equal time for all (soft and hard) modes, since the nonlinear part of the boost symmetry is constant in time, while the linearized part of the mode function is linear in time~\cite{Hui:2022dnm}.} By expanding the explicit form of $M^\mu_{\,\,\, \nu \rho}$, we can rewrite~\eqref{CRcorrelators-sub} as
\begin{align}
\frac{\partial}{\partial u}  \langle h_{ij} {\cal O}(x_1,\ldots,x_N) \rangle = \frac{\partial}{\partial u}
 \langle h_{ij} h_{k\ell} \rangle \sum_{m = 1}^N  x_m^k \frac{\partial}{\partial x_m^\ell} \left\langle{\cal O}(x_1,\ldots,x_N)\right\rangle\,.
\end{align} 
In principle, this relation can be generalized to unequal time correlators between the soft and hard modes, following Ref.~\cite{Hui:2022dnm}. We leave this generalization to future work.

\vspace{0.3cm}
\noindent {\large \bf Explicit check with~$N =2$:} Let us provide an explicit check of Eqs.~\eqref{CRcorrelators} and~\eqref{CRcorrelators-sub}, focusing on the simplest case of~$N = 2$ hard scalar modes in the presence of a planar GW propagating on Minkowski space. This requires evaluating the scalar propagator~$\langle \varphi \varphi \rangle_h$ on the planar GW background. This propagator has been constructed in Ref.~\cite{vanHaasteren:2022agf} in terms of Bessel’s functions, and it provides one of the building blocks
to studying interacting scalar field theories.

A planar GW moving in the $z$-direction with momentum $\vec k$ and frequency $\omega_g = | \vec k |\equiv k$ can be represented as
 \begin{equation}
h_{ij}(x) = h_+\epsilon_{ij}^{+}\cos\left(\omega_g u - \frac{\psi}{2}\right)
 +h_\times\epsilon_{ij}^{\times}\cos\left(\omega_g u + \frac{\psi}{2}\right)
\,,
 \label{gravitational wave: planar D=4}
 \end{equation}
with~$u = t -z$. Here,~$\psi$ denotes a constant phase difference between~$+$ and~$\times$ polarizations,~$h_+$ and~$h_\times$ are the corresponding amplitudes,
and $\epsilon_{ij}^{+}$ and $\epsilon_{ij}^{\times}$ are the two polarization tensors:
 \begin{equation}
 \epsilon_{ij}^{+}=\left(\begin{array}{ccc}
                                                 1 & 0 & 0 \cr
                                                 0 & -1 & 0 \cr
                                                  0 & 0 & 0 \cr
                                            \end{array}
                                            \right)
\,; \qquad \epsilon_{ij}^{\times}=\left(\begin{array}{ccc}
                                                 0 & 1 & 0 \cr
                                                 1 & 0 & 0 \cr
                                                  0 & 0 & 0 \cr
                                            \end{array}
                                            \right)
\,.
\label{polarization tensors}
 \end{equation}
The propagator of a scalar field of mass~$m$ on the planar GW background is given by~\cite{vanHaasteren:2022agf}\footnote{Strictly speaking, Eq.~\eqref{scalar prop full}, supplemented with~$\pm {\rm i}\epsilon$ in the time interval of the distance functions~$\Delta \bar{x}^2$, corresponds to positive/negative frequency Wightman functions~\cite{vanHaasteren:2022agf}. These can be combined as usual to construct the Feynman propagator. For simplicity, in what follows we will check the validity of the consistency relations for such Wightman functions. This is justified by the fact that our results are valid even for unequal time correlation functions~\cite{Hui:2022dnm}.}
\begin{equation}
\left\langle \varphi(x) \varphi(x') \right\rangle_h = \frac{m^2}
               {(2\pi)^2 \left[\gamma(u)\gamma(u')\right]^\frac14\sqrt{\Upsilon(u;u')}}
   \frac{K_1\left(m\sqrt{\Delta {\bar x}^2}\,\right)}
   {m\sqrt{\Delta {\bar x}^2}}
\,,
\label{scalar prop full}
\end{equation}
where~$\gamma(u)\equiv 1-  h_+^2\cos^2\left(\omega_g u - \frac{\psi}{2}\right)-h_\times^2\cos^2\left(\omega_g u + \frac{\psi}{2}\right)$ is the 
determinant of $h_{ij}$, and~$K_\nu$ is the Bessel function of the second kind. The argument of the Bessel function includes the  distance functions~$\Delta {\bar x}^2(x;x')$, given in lightcone coordinates (with~$v = t+z$) by
\begin{eqnarray}
\Delta {\bar x}^2(x;x') = 
 - \Delta u \Delta v 
  +\big(\Delta x\;\;\Delta y\big)\!\cdot\! {\mathbf\Upsilon}^{-1}
    \!\cdot\!\left(\begin{array}{c}
                         \Delta x\cr 
                        \Delta y\cr
                       \end{array}
                \right)
\,,
\label{distance function gen}
\end{eqnarray}
where~$\Delta$ indicates the difference between the coordinates. 
The matrix $\Upsilon$ is called the deformation matrix. For a polarized GW, its components are given by~\cite{vanHaasteren:2022agf} (we define $c_\psi,s_\psi = \cos \psi, \sin \psi$)        
\begin{eqnarray}
{\Upsilon}_{{}_{xx}\atop{}^{yy}} \!\!&=&\!\! 
 \frac{1}{\omega_g\Delta u \sqrt{1-\big(h_+^2 + h_\times^2\big)+ h_+^2 h_\times^2s_{\psi}^2} }
 \left\{\! \text{arctan}\left( \frac{h_\times^2 s_{\psi} c_{\psi} 
  + \big(1-h_\times^2 s_{\psi}^2\big)\tan\left(\omega_g u -\frac{\psi}{2}\right) }{\sqrt{1-\big(h_+^2 + h_\times^2\big)+ h_+^2 h_\times^2s_{\psi}^2}}\right) 
\right.
\nonumber \\ 
%&& \hskip -.2cm
&& \left. \mp  \, {\rm i}\frac{\big(1\!-\!h_\times^2 s_{\psi}^2\big)h_+
       \text{arctanh}[g(u)]}
{2\sqrt{h_+^2 \!+\! h_\times^2 \!-\! h_\times^2 s_{\psi}^2 
    \big(2\!+\!h_+^2\!-\! h_\times^2\big)\!-\! 2 {\rm i}h_\times^2 c_{\psi} s_{\psi}
     \sqrt{1\!-\!\big(h_+^2 \!+\! h_\times^2\big)\!+\! h_+^2 h_\times^2 s_{\psi}^2}}}
\!+{\rm c.c.} - (u\rightarrow u')\! \right\} ;
\nonumber\\
{\Upsilon}_{xy} \!\!&=&\!\!
 \frac{1}{\omega_g\Delta u \sqrt{1-\big(h_+^2 + h_\times^2\big)+ h_+^2 h_\times^2s_{\psi}^2} }
 \nonumber\\
\!\!&\times&\!\!\!\!
 \left\{  -\frac{\left[{\rm i}c_{\psi} 
   \!+\!   s_{\psi}\sqrt{1\!-\!\big(h_+^2 \!+\! h_\times^2\big)
\!+\! h_+^2 h_\times^2 s_{\psi}^2} \,
\right]h_\times
       \text{arctanh}[g(u)]}
{2\sqrt{h_+^2 \!+\! h_\times^2 \!-\! h_\times^2 s_{\psi}^2 
    \big(2\!+\!h_+^2\!-\! h_\times^2\big)\!-\! 2 {\rm i}h_\times^2 c_{\psi} s_{\psi}
     \sqrt{1\!-\!\big(h_+^2 \!+\! h_\times^2\big)\!+\! h_+^2 h_\times^2 s_{\psi}^2}}}
\!+{\rm c.c.}- (u\rightarrow u')\! \right\}
, \qquad\; \nonumber\\
\label{deform matrix comps}
\end{eqnarray}
with 
\begin{equation}
	g(u)\equiv\frac{\big(1\!-\!h_\times s_{\psi}^2\big)
%	\cos(\omega_g u \!-\! \psi/2) 
\cos\left(\omega_g u - \frac{\psi}{2}\right) \!+\!\left[{\rm i}\sqrt{1\!-\!\big(h_+^2 \!+\! h_\times^2\big)
	\!+\! h_+^2 h_\times^2 s_{\psi}^2}\!-\!h_\times^2 c_{\psi} s_{\psi}\right]
	\sin\left(\omega_g u \!-\! \frac{\psi}{2}\right)}
	{\sqrt{h_+^2 \!+\! h_\times^2 \!-\! h_\times^2s_{\psi}^2 
	\big(2\!+\!h_+^2\!-\! h_\times^2\big)
	\!-\!2 {\rm i}h_\times^2 c_{\psi} s_{\psi}
	\sqrt{1\!-\!\big(h_+^2 \!+\! h_\times^2\big)
	\!+\! h_+^2 h_\times^2s_{\psi}^2}}}\,.
\end{equation} 
The inverse and determinant of the deformation matrix are then given by
\begin{align}
{\mathbf\Upsilon}^{-1}(u;u') & =  \frac{1}{\Upsilon(u;u')}
               \left(\!\!\begin{array}{cc}
                         \Upsilon_{yy} & - \Upsilon_{xy}\cr 
                      - \Upsilon_{xy} &  \Upsilon_{xx} \cr
                       \end{array}
                \!\!\right)\,; \qquad
\Upsilon(u;u') \equiv   {\rm det} \left[{\mathbf\Upsilon}\right]
   = \Upsilon_{xx}\Upsilon_{yy}-\Upsilon_{xy}^2
\,.
\end{align}

\vspace{0.2cm}
\noindent
For our purposes, we only need the scalar propagator in Eq.~\eqref{scalar prop full} up to linear order in the tensor mode.
To zeroth order in~$h$, we should recover the propagator on the Minkowski background. Indeed, it is easy to see that~$\lim_{h \to 0} {\Upsilon}_{{}_{xx}\atop{}^{yy}}  = 1$,~$\lim_{h \to 0} \Upsilon_{xy} = 0$, such that~$\lim_{h \to 0} \Upsilon(u;u') =1$. It follows that the distance function~\eqref{distance function gen} reduces to the known Minkowskian result,
\begin{align}
\Delta {\bar x}_0^2(x;x') =  -\Delta u \Delta v + 
\Delta x^2 +  \Delta y^2 \,,
% -\Delta t^2 + \Delta x^2 + \Delta y^2 + \Delta z^2\,.
\label{distance zeroth order}
\end{align}
such that the scalar propagator~\eqref{scalar prop full} matches the flat-space result:
\begin{equation}
\left\langle \varphi(x) \varphi(x') \right\rangle = \lim_{h \to 0}\left\langle \varphi(x) \varphi(x') \right\rangle_h = \frac{m^2}
               {(2\pi)^2}
   \frac{K_1\left(m\sqrt{\Delta {\bar x}_0^2}\right)}
   {m\sqrt{\Delta {\bar x}_0^2}}
\,.
\label{flat space scalar correlator}
\end{equation}

At first order in the tensor amplitude, the deformation matrix components~\eqref{deform matrix comps} reduce to
\begin{align}
{\Upsilon}_{{}_{xx}\atop{}^{yy}} & = 1 \mp \frac{2 h_+ \sin \left(\frac{1}{2} \omega_g (u-u^\prime)\right) \cos \left(\frac{1}{2}(\omega_g (u+u^\prime)-\psi)\right)}{\omega_g (u-u^\prime)}\,; \nonumber \\
{\Upsilon}_{xy} & = - \frac{ 2h_\times \sin \left(\frac{1}{2} \omega_g (u-u^\prime)\right) \cos \left(\frac{1}{2} (\omega_g (u+u^\prime)+\psi )\right)}{\omega_g (u-u^\prime)}\,,
\end{align}
with the determinant $\Upsilon(u;u') = 1 + {\cal O}\big(h_{+,\times}^2\big)$.
We stress that these equations match the first-order expansion of the more general integral terms shown in Ref.~\cite{vanHaasteren:2022agf}. Thus the distance function~\eqref{distance function gen} takes the form~$\Delta {\bar x}^2(x;x') = \Delta {\bar x}_0^2(x;x') + \Delta {\bar x}_1^2(x;x')$, where~$\Delta {\bar x}_1^2(x;x')$ is a first-order correction given by  
\begin{align}
\Delta {\bar x}_1^2(x;x')  & = \frac{1}{\omega_g\Delta u} \left[
\sin \left(\frac{\psi }{2}\right) \Big(h_+ \big(\Delta x^2-\Delta y^2\big) -2h_\times \Delta x \Delta y \Big) \Big(\cos (\omega_g u^\prime )-\cos (\omega_g u )\Big) \right. \nonumber \\
& \left.~~~~~~~~~\; -\cos \left(\frac{\psi }{2}\right) \Big(h_+ \big(\Delta x^2-\Delta y^2\big) + 2h_\times \Delta x \Delta y \Big) \Big(\sin (\omega_g u^\prime)-\sin (\omega_g u )\Big) \right] \,. 
\end{align}
Substituting these results into Eq.~\eqref{scalar prop full}, and using the fact that~$\gamma(u) = 1 + {\cal O}\big(h_{+,\times}^2\big)$, we obtain
\begin{align}
\left\langle \varphi(x) \varphi(x') \right\rangle_h & \simeq \frac{m^2}{(2\pi)^2}\frac{K_1\left(m\sqrt{\Delta {\bar x}_0^2}\right)} {m\sqrt{\Delta {\bar x}_0^2}}   \nonumber \\
   & -  \frac{m^3}{(2\pi)^2}\frac{K_2\left(m\sqrt{\Delta {\bar x}_0^2}\right)} {m \Delta {\bar x}_0^2}
\frac{\sin \left( \frac{\omega_g \Delta u}{2} \right)}{\omega_g \Delta u} \nonumber \\
 & \times \left[h_+ \big(\Delta x^2-\Delta y^2\big) \cos \left(\frac{1}{2} ( \omega_g u^\prime + \omega_g u -\psi )\right) + 2 h_\times \Delta x \Delta y  \cos \left(\frac{1}{2} ( \omega_g u^\prime +  \omega_g u+\psi )\right) \right] 
   \,.
\label{scalar prop linear}
\end{align}
This is the desired scalar propagator, to linear order in the planar GW amplitude.

A long constant mode, which mimics the presence of a memory term, is achieved in the soft limit~$\omega_g \to 0$, such that Eq.~\eqref{gravitational wave: planar D=4} gives 
 \begin{align}
\bar{h}_{ij} \equiv \lim_{\omega_g \to 0} h_{ij}(x) & \simeq \left( h_+\epsilon_{ij}^{+}
 +h_\times\epsilon_{ij}^{\times}\right) \cos \left(\frac{\psi}{2}\right)  + \left( h_+\epsilon_{ij}^{+}
 -h_\times\epsilon_{ij}^{\times}\right) u \, \omega_g \sin \left(\frac{\psi}{2}\right) + \mathcal{O}(\omega_g^2) \nonumber \\
 & \equiv \frac{1}{\bar{R}} \left[A_{ij} + B_{ij} u \right] + \mathcal{O}(\omega_g^2)\,,
\label{bar h explicit}
 \end{align}
 where in the last step we have identified the modes $A_{ij}$ and $B_{ij}$ in analogy with Eq.~\eqref{hij TT local}.
In this limit the propagator~\eqref{scalar prop linear} reduces to
 \begin{align}
 \lim_{\omega_g \to 0}\left\langle \varphi(x) \varphi(x') \right\rangle_h  & \simeq \frac{m^2}{(2\pi)^2}\frac{K_1\left(m\sqrt{\Delta {\bar x}_0^2}\right)} {m\sqrt{\Delta {\bar x}_0^2}}   \nonumber \\
    & -  \frac{m^3}{2(2\pi)^2}\frac{K_2\left(m\sqrt{\Delta {\bar x}_0^2}\right)} {m \Delta {\bar x}_0^2} 
   \Big[h_+ \big(\Delta x^2-\Delta y^2\big)  + 2 h_\times \Delta x \Delta y \Big]\cos \left(\frac{\psi}{2}\right) \nonumber \\
   & -  \frac{m^3}{2(2\pi)^2}\frac{K_2\left(m\sqrt{\Delta {\bar x}_0^2}\right)} {m \Delta {\bar x}_0^2} 
   \Big[h_+ \big(\Delta x^2-\Delta y^2\big)  - 2 h_\times \Delta x \Delta y \Big] \omega_g u
   \sin \left(\frac{\psi}{2}\right) \,.
\end{align}
The left-hand side of the consistency relations~\eqref{CRcorrelators} and~\eqref{CRcorrelators-sub} then reads
\begin{align}
\label{lhsscalar}
 & \lim_{\omega_g \to 0}\left\langle h_{ij} \left\langle \varphi(x) \varphi(x') \right\rangle_h \right\rangle  \nonumber \\
% &=  -  \frac{m^3}
 %             {2(2\pi)^2}\frac{K_2\left(m\sqrt{\Delta {\bar x}_0^2}\right)} {m\Delta {\bar x}_0^2}
 %  \Bigg[\epsilon_{ij}^+\vev{h_+ h_+} \big(\Delta x^2-\Delta y^2\big) + \epsilon_{ij}^\times \vev{h_\times h_\times} 2 \Delta x \Delta y   
%   \Bigg]\cos^2 \left(\frac{\psi}{2}\right) \nonumber \\
%  & -  \frac{m^3}
%              {2(2\pi)^2}\frac{K_2\left(m\sqrt{\Delta {\bar x}_0^2}\right)} {m\Delta {\bar x}_0^2}
%   \Bigg[\epsilon_{ij}^+ \vev{h_+ h_+} \big(\Delta x^2-\Delta y^2\big) -\epsilon_{ij}^\times \vev{h_\times h_\times} 2 \Delta x \Delta y   
%   \Bigg]\left(u_i + \frac{u+u'}{2}\right) \omega_g
%   \frac{\sin \psi}{2} \nonumber \\
&= -  \frac{m^3}
              {2(2\pi)^2}\frac{K_2\left(m\sqrt{\Delta {\bar x}_0^2}\right)} {m\Delta {\bar x}_0^2}
   \Bigg[\epsilon_{ij}^+\vev{h_+ h_+} \big(\Delta x^2-\Delta y^2\big) + \epsilon_{ij}^\times \vev{h_\times h_\times} 2 \Delta x \Delta y   
   \Bigg]\cos^2 \left(\frac{\psi}{2}\right) \nonumber \\
  &   -  \frac{m^3}
              {(2\pi)^2}\frac{K_2\left(m\sqrt{\Delta {\bar x}_0^2}\right)} {m\Delta {\bar x}_0^2}
   \Bigg[\epsilon_{ij}^+ \vev{h_+ h_+} \big(\Delta x^2-\Delta y^2\big) -\epsilon_{ij}^\times \vev{h_\times h_\times} 2 \Delta x \Delta y   
   \Bigg]u \, \omega_g
   \frac{\sin \psi}{2}\,,
\end{align}
where we have evaluated all fields at equal time~$u$.
Meanwhile, the right-hand side of the consistency relations is given by 
\begin{align}
& \lim_{\omega_g \to 0}  \vev{h_{ij} \left[M_{mn} \left(x^n \frac{\partial}{\partial x^m} + x^{\prime n} \frac{\partial}{\partial x^{\prime m}} \right) +  M^\mu_{\,\,\, \nu \rho} \left(x^\nu x^\rho \frac{\partial}{\partial x^\mu} + x^{\prime \nu} x^{\prime \rho} \frac{\partial}{\partial x^{\prime \mu}} \right) \right] } \left\langle \varphi(x) \varphi(x') \right\rangle \nonumber \\
& = \lim_{\omega_g \to 0} \left\langle
\left[\frac{A_{ij}}{\bar{R}} + \frac{B_{ij}}{\bar{R}} u_i \right] \left[\frac{A_{mn}}{2\bar{R}} \left(x^n \frac{\partial}{\partial x^m} + x^{\prime n} \frac{\partial}{\partial x^{\prime m}} \right) 
+ \frac{B_{mn}}{2\bar{R}} \left(x^n u \frac{\partial}{\partial x^m} + x^{\prime n} u^{\prime} \frac{\partial}{\partial x^{\prime m}} \right)  \right. \right. \nonumber \\
& \left. \left. + \frac{B_{nk}}{4\bar{R}} \left( x^n x^k  \frac{\partial}{\partial v} + x^{\prime n} x^{\prime k}  \frac{\partial}{\partial v^{\prime}} \right) \right] \right\rangle \left\langle \varphi(x) \varphi(x') \right\rangle  \,, 
\label{interm eqn with d/dv}
\end{align}  
where in the second line we have expanded the soft mode $h_{ij}$ following Eq.~\eqref{bar h explicit} and  we have substituted the expressions for the matrices $M_{mn}$ and $M^\mu_{\,\,\, \nu \rho}$. Explicitly carrying out the derivatives
of the flat-space scalar propagator in Eq.~\eqref{flat space scalar correlator} and evaluating the result at equal time~$u$,\footnote{The~$\partial/\partial v$ contributions in the last line of Eq.~\eqref{interm eqn with d/dv}, which are proportional to~$\Delta u$, vanish in this case.}
we obtain
\begin{align}
& \lim_{\omega_g \to 0}  \vev{h_{ij} \left[M_{mn} \left(x^n \frac{\partial}{\partial x^m} + x^{\prime n} \frac{\partial}{\partial x^{\prime m}} \right) +  M^\mu_{\,\,\, \nu \rho} \left(x^\nu x^\rho \frac{\partial}{\partial x^\mu} + x^{\prime \nu} x^{\prime \rho} \frac{\partial}{\partial x^{\prime \mu}} \right) \right] } \left\langle \varphi(x) \varphi(x') \right\rangle \nonumber \\
   & =  -  \frac{m^3}
               {2 (2\pi)^2}\frac{K_2\left(m\sqrt{\Delta {\bar x}_0^2}\right)} {m \Delta {\bar x}_0^2 }
   \Bigg[\epsilon^+_{ij}\vev{h_+ h_+} \big(\Delta x^2-\Delta y^2\big) + \epsilon^\times_{ij} \vev{h_\times h_\times} 2 \Delta x \Delta y   
   \Bigg]\cos^2 \left(\frac{\psi}{2}\right) \nonumber \\
& - \frac{m^3}
               {(2\pi)^2}\frac{K_2\left(m\sqrt{\Delta {\bar x}_0^2}\right)} {m \Delta {\bar x}_0^2 }
   \Bigg[\epsilon^+_{ij} \vev{h_+ h_+} \big(\Delta x^2-\Delta y^2\big) -  \epsilon^\times_{ij}\vev{ h_\times h_\times} 2 \Delta x \Delta y   
   \Bigg] u \, \omega_g \frac{\sin \psi}{2}\,, 
\end{align}
where we have used the polarization tensors given in Eq.~\eqref{polarization tensors}. This result precisely matches Eq.~\eqref{lhsscalar}, thus proving the validity of the leading and subleading consistency relations for the scalar propagator.

\section{freely falling detectors}

\begin{figure}[t]
    \centering
    \includegraphics[scale=0.4]{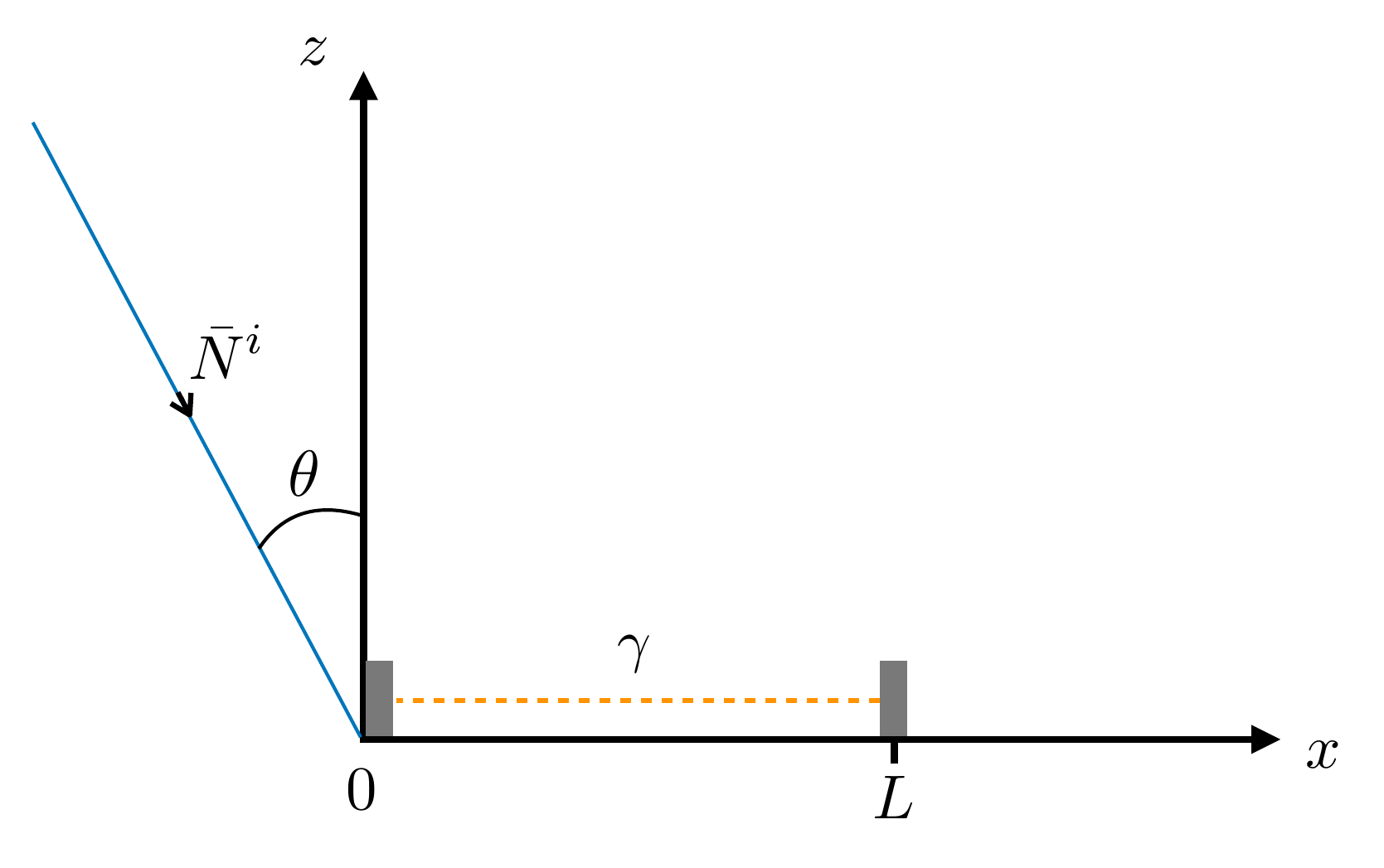}
    \caption{Simplified geometry of a GW detector. The GW (in blue) is incident along the~$\bar{N}_i$ direction, which lies in the $x-z$ plane. The arm of the laser interferometer consists of two freely falling mirrors (thick gray line), separated by a distance~$L$ along the~$x$ axis. Photons (yellow dashed line) travel from $x = 0$ to~$x = L$, bounce back from~$x = L$ and return to the origin.}
    \label{fig:Laser}
\end{figure}

It is an elementary yet instructive exercise to see how the effect of a GW given by the long mode~\eqref{hij TT local} on physical observables can be fully removed by a change of coordinates. For this purpose, consider the classic problem of photons bouncing between two freely falling mirrors, which constitute one arm of a laser interferometer. See Fig.~\ref{fig:Laser} for  an illustration. The arm is assumed to lie along the~$x$ axis, with mirrors located at~$x = 0$ and~$x = L$.

Consider an incident planar GW with ``+" polarization, $h_+(u)$, propagating in a direction~$\bar{N}_i$. Without loss of generality,~$\bar{N}_i$ can be taken to lie in the~$x-z$ plane, with polar angle~$\theta$. 
In TT gauge, the induced metric along the~$x$ axis is 
\be
{\rm d}s^2_x = -{\rm d}t^2 + \Big(1 + h_+(t-x\sin\theta) \cdot \cos^2\theta\Big) {\rm d}x^2\,.
\ee
A convenient aspect of TT gauge is that a freely falling test mass, originally at fixed coordinates, remains at fixed coordinates in the presence of a GW~\cite{Arnowitt:1962hi}.

The observable of interest is the time taken by a photon to travel from~$x = 0$, reflect off the mirror at~$x = L$, and come back to the origin. It is easy to show that the round-trip time is given by~\cite{Maggiore:2007ulw}
\be
T = t_0 + 2L + \frac{1}{2} \cos^2\theta \int_0^L {\rm d} x \bigg[h_+\big(t_0 + x(1-\sin\theta)\big)  + h_+\big(t_0 + 2L - x(1+\sin\theta)\big)\bigg]\,,
\label{round trip}
\ee
where~$t_0$ is the time at which the GW is detected. The first terms,~$t_0 + 2L$, give the round-trip time in flat space; the remainder is the perturbation due to the GW. Now let us consider a long mode, given by the leading terms in Eq.~\eqref{hij TT local}:
\be
h_+ (u) = \frac{1}{\bar{R}} \big( A + B u \big)\,,
\label{h+ long}
\ee
where~$A$ and~$B$ are constants, and~$u = t - \bar{N}_ix^i = t - x\sin\theta - z\cos\theta$, as before. Substituting into~\eqref{round trip}, the round-trip time evaluates to
\be
T =  t_0 + 2L + \frac{L}{\bar{R}} \cos^2\theta \Big(A + Bt_0\Big) +  \frac{B L^2}{\bar{R}}\cos^2\theta \left( 1 -  \frac{1}{2}\sin\theta\right)\,.
\label{round trip long}
\ee
The long mode~\eqref{h+ long} can be removed by the spatial and time diffeomorphisms given by~\eqref{diff in terms of u} and~\eqref{time diff in terms of u} respectively.
In the situation of interest, the coordinate transformation is
\begin{align}
\nonumber
\tilde{t} &= t + \frac{1}{4\bar{R}} B\cos^2\theta \, x^2 \,;\\
\tilde{x} &= x + \frac{1}{2\bar{R}} \cos^2\theta \big(A + B t\big) x - \frac{1}{4\bar{R}} B \cos^2\theta \sin\theta \, x^2\,.
\label{diff bouncing}
\end{align}
Since there is no GW in the new coordinate system, the round-trip time is simply given by
\be
T =  t_0 + 2\tilde{d} \,,
\label{round trip long 2}
\ee
where~$\tilde{d}$ is the spatial distance traveled. To calculate the latter, note that the first mirror (originally at~$x = 0$) remains at~$\tilde{x} = 0$. The impact of the diffeomorphism~\eqref{diff bouncing} on the second mirror is twofold. First, the spatial diffeomorphism at time~$t_0$ implies a stretching of the coordinate distance:
\be
\tilde{L} = L + \frac{1}{2 \bar{R}} L\cos^2\theta \big(A + B t_0\big) - \frac{1}{4 \bar{R}} BL^2 \cos^2\theta \sin\theta  \,.
\ee
Second, in the new coordinate system, the second mirror is now moving with velocity~$\partial \tilde{x}/ \partial t  = \frac{1}{2 \bar{R}} B L \cos^2\theta$, which means that the photon must travel an additional distance
\be
\Delta \tilde{x} \simeq \frac{1}{2 \bar{R}}B L^2 \cos^2\theta
\ee
to reach it. Thus the total spatial distance is the sum of~$\tilde{L}$ and~$\Delta\tilde{x}$, which gives
\be
\tilde{d} = L + \frac{L}{2\bar{R}} \cos^2\theta \big(A + B t_0\big) + \frac{B L^2}{2\bar{R}} \cos^2\theta \left( 1 - \frac{1}{2} \sin\theta\right)\,.
\ee
Thus the round-trip time~\eqref{round trip long 2} agrees with~\eqref{round trip long}, which confirms that this physical observable can be similarly described through a coordinate transformation that takes into account the long mode.

\section{Conclusions}
Gravitational memory effects describe a lasting change in the GW strain associated with the evolution of a coalescing binary system. It has been shown to impact the relative separation between two freely falling detectors before and after the GW event. Memory effects are intricately related to the BMS symmetry group of asymptotically flat space-times, and they were proven to be equivalent to transitions between two distinct asymptotic BMS frames connected by a supertranslation.

Since GW measurements for experiments in free fall are usually described in the TT frame, it is natural to inquire about the form of the residual coordinate transformations one could perform, in the local frame around the detector, to describe the memory effect. In this work we have identified the large residual diffeomorphisms in TT gauge, which describe gravitational memory. For instance, the constant TT mode, describing the shift induced by memory effects, corresponds to an anisotropic (volume-preserving) spatial rescaling. Similarly, the constant velocity kick induced by memory term can be removed in the TT gauge by performing a time-dependent anisotropic spatial rescaling,
together with a spatial diffeomorphism that removes a homogeneous acceleration, familiar from the equivalence principle. Importantly, we have shown that these residual diffeomorphisms are precisely equivalent to BMS transformations, together with suitable compensating diffeomorphisms to restore TT gauge. This bridges the gap between the widely recognized large diffeomorphism familiar to cosmologists and the asymptotic BMS symmetries.

Starting from the Ward identities for the local residual diffeomorphisms, we then derived the corresponding soft theorems, both for scattering amplitudes and for equal-time correlation functions. Using the explicit form of the diffeomorphisms, made out of a linear and quadratic piece in the spatial coordinates around the detector, we first determined the tree-level leading and subleading soft theorems for scattering amplitudes. They relate the scattering amplitude/in-in correlator with~$N$ hard modes with the~$N+1$ one involving an extra soft graviton mode, which represents the asymptotic zero-frequency limit of the memory term. For equal-time correlation functions, we similarly derived the explicit soft theorems for hard scalar field modes and checked their validity with the simple example of a planar GW.

Lastly, as an illustrative physical check, we considered a simplified model of a laser interferometer, where a photon travels along an arm between two mirrors, and showed that the effect of a long memory mode on the round-trip time can be mimicked by the action of the residual diffeomorphism on the mirrors. This computation provides an instructive example of the action of the residual diffeomorphism on GW observables.

Our work provides a further step in understanding the interplay between GW memory, soft theorems and symmetries, and can be extended in several directions. An immediate possibility is to generalize our soft theorems to hard tensor modes. As mentioned earlier, these would involve a slightly more complicated structure, because of the linear transformation required to preserve TT gauge. A second direction for further investigation is to extend our framework to higher-order memory effects, such as the spin and center of mass memory effects. It has been argued that these should arise when considering the noninertial motion of detectors, thereby generalizing the geodesic deviation equation to an initially accelerated motion~\cite{Flanagan:2019ezo,Siddhant:2024nft}. 
This would represent a further step  to deepen the understanding of the interplay with soft theorems and asymptotic symmetries in the infrared triangle. Finally, it would be interesting to extend the computation of soft theorems for scattering amplitudes including loop corrections, to highlight the deep connection with tails-of-memory effects and superrotation symmetries.
We plan to investigate these directions in future work.

\subsubsection*{Acknowledgments}
We thank D. Nichols for interesting discussions, and an anonymous referee for valuable comments regarding loop-order corrections to the soft theorems for scattering amplitudes.
V.DL. is supported by funds provided by the Center for Particle Cosmology at the University of Pennsylvania. 
The work of J.K. is supported in part by the DOE (HEP) Award DE-SC0013528.
The work of S.W. is supported by APRC-CityU New Research Initiatives/Infrastructure Support from Central.

\begin{appendices}

\section{Path integral derivation of consistency relations for correlators}
\renewcommand{\theequation}{A.\arabic{equation}}
\setcounter{equation}{0}
\label{AppA}

In this Appendix we schematically review the path integral approach to derive the in-in consistency relations or soft theorems for correlation functions~\cite{Hui:2018cag}. As a starting point, let us consider a general operator $\hat{{\cal O}}(x_1,\dots, x_N)$, built from fundamental fields $\Phi$ evaluated at the same late time $t$. Its in-in correlator  can be written as
\begin{align}
   & \vev{\Omega |\hat{{\cal O}}(x_1,\dots, x_N) | \Omega} \nonumber \\
   & = \int \left[ {\cal D} \Phi^+_0 {\cal D} \Phi^+ {\cal D}\Phi^-_0 {\cal D}\Phi^- \right]  \vev{\Omega| \Phi^+_0} \vev{\Phi^+_0|\Phi^+} \vev{\Phi^+|\hat{{\cal O}}(x_1,\dots, x_N)|\Phi^-} \vev{\Phi^-|\Phi^-_0}\vev{\Phi^-_0|\Omega} \nonumber \\
   & = \int \left[{\cal D} \Phi \right] \Psi^{\dagger}[\Phi]\Psi[\Phi] {\cal O}(x_1,\dots, x_N)\,,  
\end{align}
which is based on the  double path integral (Schwinger-Keldysh) representation~\cite{Schwinger:1960qe, Keldysh:1964ud, Feynman:1963fq, Chou:1984es}. The wave functional $\Psi[\Phi]$ can be constructed, from some initial vacuum state $|\Omega \rangle$,  by inserting a complete set of field eigenstates as
\begin{equation}
    \Psi[\Phi] =  \raisebox{.09cm}{$\underset{\substack{{\scriptscriptstyle \Phi(t)=\Phi} \\ {\scriptscriptstyle \Phi(-\infty) = \Phi_0}}}{\displaystyle\int}$}  \left[{\cal D}\Phi _0\right] \left[{\cal D}\Phi \right] {\rm e}^{{\rm i} S[\Phi]} \Psi_0\,,  
    %\qquad \Psi_0=\vev{\Phi_0|\Omega}\,,
\end{equation}
with~$\Psi_0=\vev{\Phi_0|\Omega}$, where the path integral sums over all possible field configurations subject to the boundary condition at early time~$\Phi(t_0 \to -\infty) = \Phi_0$, and~$\Phi (t) = \Phi$. The vacuum wave functional~$\Psi_0 = \vev{\Phi_0|\Omega}$ is evaluated at the infinite past. Assuming that interactions are adiabatically switched on, it can be approximated using the free theory Gaussian wave functional as~\cite{Weinberg:1995mt}
\begin{align}
    \Psi_0[\Phi_0] \propto \exp\left[-\frac{1}{2}\int\frac{\rd^3 k}{(2\pi)^3}{\cal E}_0(k)\Phi_0 \big(\vec{k} \big)\Phi_0\big(-\vec{k}\,\big) \right]\,,
\end{align}
in terms of some kernel~${\cal E}_0$, where we have assumed rotational symmetry.

Consider an abstract symmetry acting on~$\Phi$. Under an arbitrary field transformation~$\Phi \to \Phi +  \delta \Phi $, assuming that the integration measure~$\left[{\cal D} \Phi \right] $ is invariant, the path integral evaluates to the same result, such that
\begin{align} 
\label{eqn:pathinvariation}
    0 =  \int \left[{\cal D} \Phi \right] \Psi^{\dagger}[\Phi]\Psi[\Phi] \delta {\cal O}(x_1,\dots, x_N) +  \int \left[{\cal D} \Phi \right] \Big( \delta \Psi^{\dagger}[\Phi] \Psi[\Phi] +\Psi^{\dagger}[\Phi] \delta \Psi[\Phi] \Big){\cal O}(x_1,\dots, x_N)\,,   
\end{align}
where $\delta {\cal O}$ is the transformation rule for the operator ${\cal O}$ inherited from its dependence on $\Phi$.

Focusing on transformations $\delta \Phi$ that are (spontaneously broken) symmetries of the theory, the variation of the wave functional  $\delta \Psi[\Phi]$ can be written as 
\begin{align}
    \delta \Psi[\Phi] = \Psi[\Phi+ \delta \Phi]-\Psi[\Phi]  =  \raisebox{.09cm}{$\underset{\substack{{\scriptscriptstyle \Phi(t)=\Phi} \\ {\scriptscriptstyle \Phi(-\infty) = \Phi_0}}}{\displaystyle\int}$}  \left[{\cal D}\Phi _0\right] \left[{\cal D}\Phi \right] {\rm e}^{{\rm i} S[\Phi]} \delta \Psi_0 \;, 
\end{align}
where we have ignored spatial boundary terms in the action, and used the fact that 
\begin{align}
    \left\langle\Phi(t) + \delta \Phi(t)   |\Phi(t_0) +\delta \Phi(t_0)\right\rangle =  \vev{\Phi(t)  |\Phi(t_0) }\,,
\label{app titf}
\end{align}
as long as $\delta \Phi$ is a symmetry of the theory and does not generate time boundary terms.\footnote{This is valid provided the symmetries being considered are purely spatial. In the case of ``time symmetries", care must be taken in handling time boundary terms~\cite{Hui:2022dnm}. This issue should be considered when studying the unequal time consistency relations for the linear gradient mode discussed in the main text.} Let us consider symmetries that act nonlinearly on the fields,~$\delta \Phi = \delta_\text{\tiny NL}\Phi + \delta_\text{\tiny L} \Phi$, {\it i.e.}, that consist of both a nonlinear (not proportional to the field itself) and a linear piece. Assuming that the nonlinear part does not vanish at spatial infinity, it can be written in momentum space as 
\begin{align}
\label{Dq}
    \delta_\text{\tiny NL}\Phi \left(\vec{q} \right)  = (2\pi)^3\delta^{(3)}(\vec{q}) {\cal D}_{-\vec{q}}\,, 
\end{align}
where~${\cal D}_{-\vec{q}}$ is some derivative operator depending on~$\vec{q}$. Therefore, the variation of the initial wavefunctional~$\delta \Psi_0$ can be easily computed to be
\begin{align}
    \delta \Psi_0[\Phi_0] = - {\cal D}_{\vec q} \Big[{\cal E}_0(q)\Phi_0\left(\vec{q} \right)  \Big] \Big\vert_{{\vec q} = 0} \Psi_0[\Phi_0]\,,
\end{align}
where we have ignored the $\delta_\text{\tiny L} \Phi$ pieces in $\delta \Psi_0[\Phi_0]$ following the prescription of Ref.~\cite{Hui:2018cag}. Collecting terms in Eq.~\eqref{eqn:pathinvariation}, we arrive at the following identity
\begin{align}
\label{unequal}
    {\cal D}_{\vec q} \left. \left[{\cal E}_0(q)  \vev{ {\cal O}(\vec{k}_1,\dots, \vec{k}_N)   \Phi_0 \left(\vec{q} \right) }_{\rm c}^\prime + {\rm h.c.}  \right] \right|_{\vec q =0} = \vev{\delta_\text{\tiny L}{\cal O}(\vec{k}_1,\dots, \vec{k}_N)   }_{\rm c}^\prime\,,
\end{align}
where the prime denotes a correlation function with the delta function removed, and where on the right-hand side we have removed terms associated with~$\delta_\text{\tiny NL}{\cal O}$ by focusing on the connected part of the correlators, as shown in Appendix~C of Ref.~\cite{Hinterbichler:2013dpa}. Notice that the correlator on the left-hand side is an unequal time correlator since, while the operator~$\cal{O}$ is evaluated at the final time $t$, the soft mode $\Phi_0 \big(\vec{q} \big)$ is inserted at the initial time~$t_0 \to -\infty$. As long as the physical mode condition~\cite{Hui:2018cag} is satisfied ({\it i.e.}, that the nonlinear part of the transformation has the same time dependence as the zero-momentum limit of the field), one can promote this identity to an equal time identity 
\begin{align}
\label{eqn:eqTidentityQ}
     {\cal D}_{\vec q} \left. \frac{\vev{ {\cal O}\big(\vec{k}_1,\dots, \vec{k}_N\big)   \Phi \left(\vec{q}\, \right) }_{\rm c}^\prime }{\vev{ \Phi \left(\vec{q}\, \right) \Phi \left(-\vec{q}\, \right)}' }\right|_{\vec q =0} = \vev{\delta_\text{\tiny L}{\cal O}\big(\vec{k}_1,\dots, \vec{k}_N\big) }_{\rm c}^\prime\,,
\end{align}
where the soft mode~$ \Phi \left(\vec{q}\,\right)$ is now inserted at the same time as~${\cal O}\big(\vec{k}_1,\dots, \vec{k}_N\big)$, and the kernel has been rewritten as~$\mathcal{E}_0(q)^{-1} = \vev{ \Phi \left(\vec{q}\, \right) \Phi \left(-\vec{q}\, \right)}'$. This equation describes the consistency relation for in-in correlators.

\end{appendices}
 
\bibliographystyle{JHEP}
\bibliography{draft.bib}

\end{document}